\documentclass[prb,showpacs,twocolumn,floats,10pt,aps,citeautoscript,longbibliography,superscriptaddress]{revtex4-2}
\usepackage{amsmath}
\usepackage{amssymb}
\usepackage{amsfonts}
\usepackage{braket}
\usepackage{xcolor}
\usepackage{appendix}
\usepackage{enumitem}
\usepackage{graphicx}
\usepackage{float}
\usepackage{tikz}
\usepackage{algorithm}
\usepackage{algpseudocode}
\usepackage{bbm}
\usepackage{mathtools}
\usepackage[section]{placeins}
\usepackage{lipsum, babel}

\RequirePackage[hyperindex,colorlinks,bookmarksnumbered, plainpages=true,pdfstartview=FitH]{hyperref}
\hypersetup{linkcolor=blue,urlcolor=blue,citecolor=blue}
\usepackage{hyperref}

\definecolor{darkgreen}{rgb}{0,0.5,0}

\begin{document}
  \title{Tailoring tensor network techniques to the quantics representation for highly inhomogeneous problems and few body problems}
  \author{Jheng-Wei Li}
  \affiliation{Universit\'e Grenoble Alpes, CEA, Grenoble INP, IRIG, Pheliqs, F-38000 Grenoble, France}
  \author{Nicolas Jolly}
  \affiliation{Universit\'e Grenoble Alpes, CEA, Grenoble INP, IRIG, Pheliqs, F-38000 Grenoble, France}
  \author{Xavier Waintal}
  \affiliation{Universit\'e Grenoble Alpes, CEA, Grenoble INP, IRIG, Pheliqs, F-38000 Grenoble, France}
  
\begin{abstract}
Tensor network techniques are becoming increasingly popular tools to solve
partial differential equations within the so-called quantics representation.
Their popularity stems from the fact that their spatial resolution depends only logarithmically on the number of grid points, making them very tempting approaches in situations where two or more characteristic length scales are vastly different.
A first generation of technique used ``out-of-the-box'' algorithms of the tensor network toolkit (e.g. the celebrated Density Matrix Product State (DMRG) algorithm) to solve these problems. These techniques were designed for situations (e.g. quantum magnetism) where the different degrees of freedom (e.g. spins) play equivalent roles.
In the quantics representation, however, the different degrees of freedom correspond to the physics at different scales and therefore play inequivalent role. Here we show that by tailoring the tensor network algorithms to this particular case, in the spirit of the multigrid approach, we obtain faster and more robust convergence of the algorithms.
We showcase the approach on linear (Poisson equation) and eigenvalue (Schr\"odinger equation) problems in two, three and four dimensions. Our simulations involve up to $2^{80}$ grid points and would represent, we argue, a very strong challenge for conventional approaches.
\end{abstract}
\date{\today}
\maketitle

\section{Introduction}
As a general rule, large differences in magnitude between different characteristic length scales are hard to address numerically because one needs to use a discretization dictated by the smallest of these scales. This is somewhat paradoxical because the same property can be harvested in analytical works to solve the problem. For instance, if one studies the dynamics of electrons with wavelength $\lambda$ subject to scattering with a mean free path $\ell\gg\lambda$, one can integrate out the ballistic motion taking place at scales smaller than $\ell$ and derive an effective diffusion equation valid on scales much larger than $\ell$. In that case, the existence of a small parameter $\lambda/\ell\ll 1$ provides an efficient route to solve the problem; the hard case is the crossover regime $\ell \sim \lambda$ (where Anderson localization occurs) \cite{Akkermans2020}. Another example is geometrical optics which emerges when one considers mirrors and lenses of characteristic sizes $L\gg \lambda$. A last example is homogenization problems in e.g.  diffusion equation in rapidly varying media \cite{Sanchez2009}. There, one can show that an effective diffusion equation emerges at large scale (albeit with a diffusion constant that depends non-trivially on the microscopic media).
Surely, if we can take advantage of the separation of scales analytically, there should be a way to do it numerically too, if only through an algorithmic path that follows closely what the analytics do.

\subsection{The quantics representation}
An implementation of such a program is the quantics tensor train (QTT) representation which naturally expresses functions in terms of their behavior at different scales.
QTT encodes discretized functions on $2^N$ grid points using only $N$ binary degrees of freedom and an $O(N)$ memory footprint \cite{Oseledets2009, Oseledets2011a, Oseledets2012,Dolgov2012,Chen2023,Lindsey2023}. 
In recent years, QTT methods have rapidly gained attraction across multiple disciplines including classical turbulence \cite{Gourianov2022-vn,peddinti2023complete,kornev2023,holscher2024,Gourianov2024,Huelst2025}, plasma physics \cite{Ye2022}, multidimensional Fokker–Planck equations \cite{Chertkov2021a}, quantum chemistry \cite{C5CP01215E,Jolly2025}, Gross-Pitaevskii equation \cite{BouComas2025,Chen2025,Connor2025,Niedermeier2025} and quantum (many-body) physics \cite{BENNER2017,Shinaoka2023,Erpenbeck2023,Ritter2024,Murray2024,OteroFumega2024,Kiese2024,
 Rohshap2025,Kim2025,Gidi2025}. 
By representing data with only logarithmically many parameters, QTT provides a compact and powerful computational framework with practical tools to perform important operations such as differentiation, integration, Fourier transform...with an $O(N)$ complexity (see \cite{Waintal2026} for an introduction). The key point is that the problem must possess a certain separation of scales for the associated functions to be easily compressible in the QTT representation. 

\begin{figure}[!htb]
  \centering
  \includegraphics[width=1.0\linewidth,trim = 0.0in 2.5in 0.0in 0.0in,clip=true]{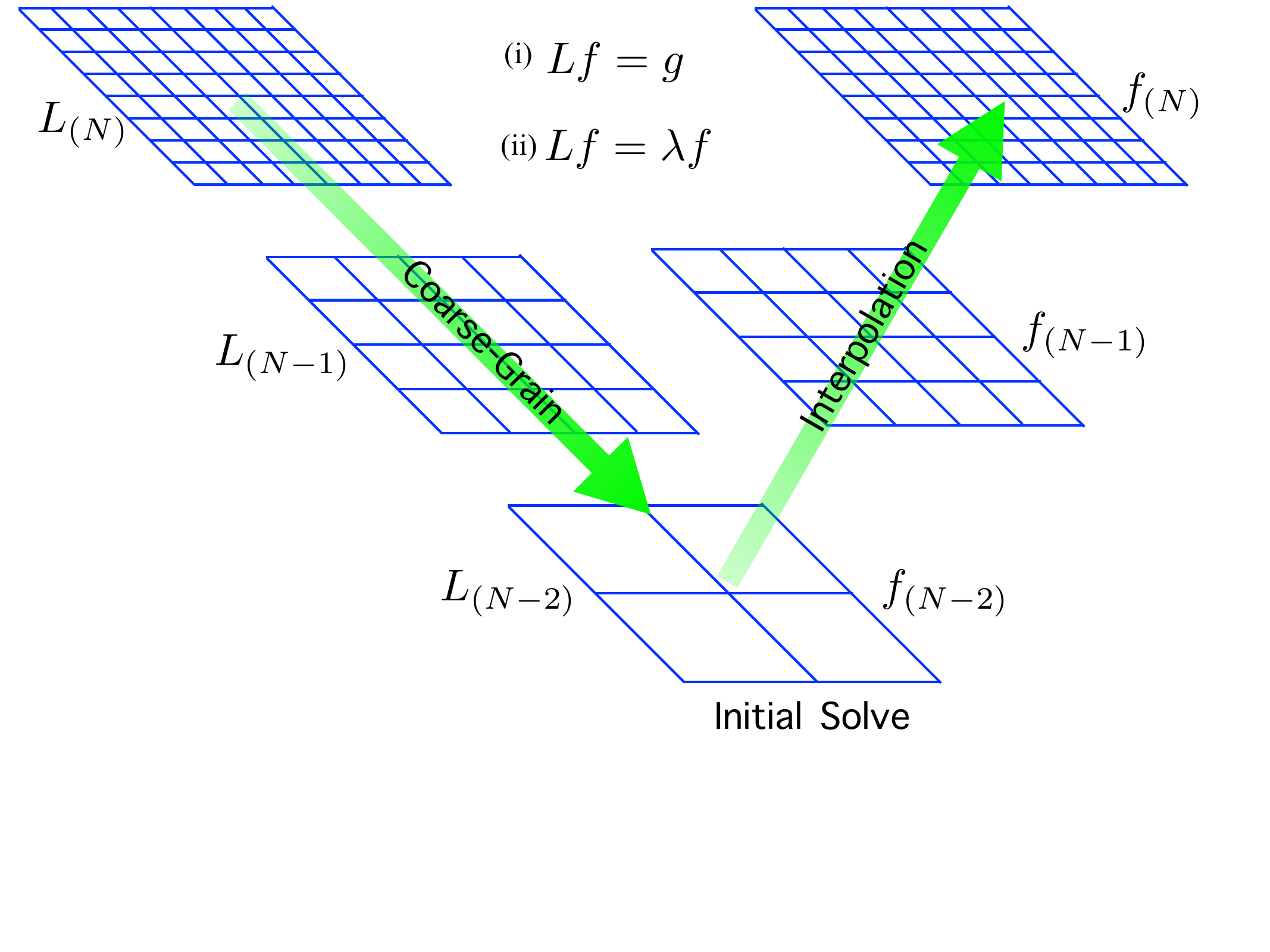}
  \includegraphics[width=1.0\linewidth,trim = 0.0in 0.0in 0.0in 0.0in,clip=true]{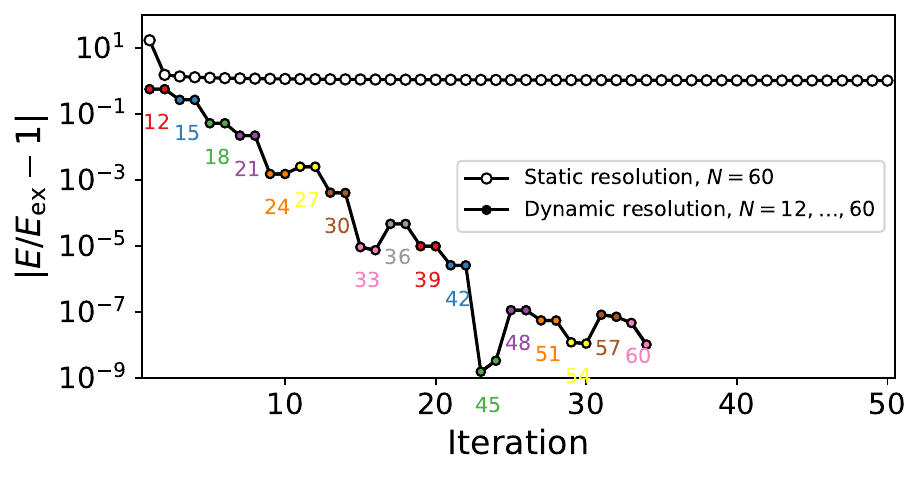}
  \caption{\textbf{Top}: The V-cycle of the multi-grid method. When descending, the problem ($L$) is coarse-grained down to a sparser grid. At the coarsest level, the problem is initially solved. Its solution then ascends by successive interpolation and is refined at every level until reaching the finest grid.
  \textbf{Bottom}: QTT calculation of the lowest eigenvalue of the hydrogen molecule ion $H_2^+$
  using plain DMRG (static resolution, $N=60$) or the multigrid DMRG approach
  (dynamical resolution $N=12\cdots 60$). The ultimate resolution of this 3D problem is  $2^{N/3}\times 2^{N/3}\times 2^{N/3} = 2^N \approx 10^{18}$ grid points for our highest resolution  $N=60$. The plot show the error on the energy as a function of the number DMRG optimization steps. In the dynamical resolution strategy, we solve at a given resolution $N$, increase $N$ to $N+3$ by interpolation, solve again...and repeat until the final resolution $N=60$ is reached. Clearly in this case, the convergence is much more robust than in the static resolution approach.}
  \label{fig:Intro}
\end{figure}

This article is concerned with the problem of scaling up QTT algorithms to  use-cases beyond the proof of principle stage. We will consider two such problems. The first is solving the Poisson equation with a highly oscillatory charge density. Here, the linear problem of the form $\Delta u=\rho$ where $\Delta$ (the discretized Laplacian) and $\rho$ (the discretized charge density) can be put, respectively, in the form of a Matrix Product Operator (MPO) and Matrix Product State (MPS) while we seek the solution $u$ as an MPS.
The second problem is a three body quantum system in four dimensions. After discretization, it will take the form of an eigenvalue problem of the form $H \Psi = E\Psi$ where $H$ is a MPO while we seek the eigenvector $\Psi$ in the form of a MPS. Traditionally, one uses solvers developed in the context of many-body physics: an Alternating Least Square (ALS) solver \cite{Holtz2012} for the linear problem and the Density Matrix Renormalization Group (DMRG) solver \cite{White1992} for the eigenvalue problem. 
However, these techniques have been designed for problems where the degrees of freedom are equivalent and may not be optimal for QTT where they are not. Here, we seek to leverage on the specific properties of QTT to speedup the convergence.

\subsection{Multi-resolution approaches for iterative solvers}

Specifically, we shall make use of the fact that QTT naturally implement the concepts needed in multigrid calculations. Let us mention that upon completion of this work, we learned of 
Ref.\cite{Lubasch2018} where this idea has already been pursued, although for different problems and smaller dimensions. The basic idea is straightforward: from a QTT at a given resolution, it is easy to average out the smaller scales to obtain a coarser QTT or to increase the resolution by interpolation to work on a finer grid. 
It follows that all the usual multigrid strategies used to improve the convergence of iterative solvers \cite{Brandt1986, Mandel1989} -- solve at a coarse level, interpolate on a finer grid, solve again etc -- are fully compatible, one could even say naturally compatible, with the QTT representation.

More precisely, iterative methods (such as Krylov based methods) are appealing for very large problems with sparse structures, because of their low computational cost (compared to direct solvers). However, when the resolution is very high, the number of iterations increases drastically. This is due to the operators that are iterated being mostly local, such that many iterations are needed for the information to propagate from one end of the system to the other. The idea of multigrid methods is relatively straightforward: one goes back and forth between different levels of resolution using a process of restriction (decreasing the resolution to a coarser grid using e.g. some sort of averaging) and prolongation (increasing the resolution to a finer grid using some sort of interpolation). Solving the problem on the different levels of resolution is expected to improve the convergence: at low resolution, the information travels the entire system within a few iterations while at high resolution, one refines the solution locally. The multigrid strategy has been developed in various contexts including for Monte Carlo integration \cite{Goodman1986,Wolfhard1994}, electronic structures \cite{Briggs1996, Heiskanen2001, Brannick2008, Dolfi2012}, and fluid dynamics \cite{Liu1998, Zhaoli2004}.

The simplest of such schemes is the so-called V-cycle algorithm which is illustrated in Fig.~\ref{fig:Intro}a. In the V-cycle, one starts with defining the problem ($L$) on the finest mesh grid, and then one iteratively maps it onto a hierarchy of coarser grids 
($\diagdown$ part of the V). When one has reached the coarsest problem, one solves it. 
Then one iteratively interpolates the solution on a finer grid and uses the result of this interpolation as the initial guess for solving  the finer grid problem ($\diagup$ part of the V).
One can implement more sophisticated multigrid patterns, known as e.g. the F-cycle or the W-cycle, with multiple back and forth between the different resolution levels to further accelerate or stabilize the convergence. In this work, we focus on the V-cycle.

To illustrate how this idea works when combined with QTT, Fig.\ref{fig:Intro} shows
an example of the behavior encountered in our numerical experiments.
Here we seek the solution of the ground state of the Schr\"odinger equation 
$-(\hbar^2/2m) \Delta\Psi +V\Psi = E\Psi$ in three dimensions where 
$V(\mathbf{r}) = -(e^2/4\pi\epsilon_0) (1/|\mathbf{r}-\mathbf{r}_A|
+ 1/|\mathbf{r}-\mathbf{r}_B|)$ is the attractive potential created by two protons situated
at positions  $\mathbf{r}_A$ and $\mathbf{r}_B$ 
(Schr\"odinger equation for the $H_2^+$ ion).
After discretization, we will use a grid of $2^{20} \times 2^{20} \times 2^{20} \sim 10^{18}$ grid points. 
We observe that the standard (two-sites) DMRG algorithm struggles to converge (open circles) while the multigrid strategy, where we alternate DMRG steps with increasing the resolution, converges quickly (filled symbols). 
The appeal for the approach is obvious in terms of robustness and speed.

This article is organized as follows: the next section \ref{sec:multigrid} introduces the necessary concepts and algorithms. Section \ref{sec:poisson} shows a first application to the Poisson equation in presence of a quickly varying charge density. Section \ref{sec:schro} shows a second application to the calculation of the ground state and excited states of the $H_2^+$ di-hydrogen ion, including its vibrational degrees of freedom.

\section{QTT optimization with dynamical resolution} 
\label{sec:multigrid}

In this section, we introduce our QTT notations (this is standard material so we refer to the literature for the proofs of the main statements). Then, starting from a given QTT that represents a function at a given resolution, we explain how to construct another QTT of the same function with finer (through interpolation) or coarser (through averaging or sampling) grids. Last, we combine these tools with ALS and DMRG solvers to construct a linear and an eigenvalue solver with better convergence properties than the plain fixed resolution approach.

\subsection{Definition of the QTT representation}

We start by defining the QTT representation. Let $f(\mathbf{x})$ be a scalar function of interest with $\mathbf{x} \in [a,b]^d$ in $d$ dimensions (e.g. an input of the partial differential equation or the sought solution). Let us first focus on the one dimensional case $d=1$. We begin by simply discretizing the interval $[a,b]$ onto an exponentially large number of $2^N$ equally separated points $x_\alpha$, 
\begin{align}
x_{\alpha} = a + \frac{b-a}{2^N} \alpha \quad \alpha \in 0, \ldots, 2^N-1.
\end{align}
We obtain a very large vector $f_\alpha \equiv f(x_\alpha)$. 
Second, we write each of these integers $\alpha$ in (its unique) binary format 
$\alpha = 2^{N-1} s_1^{(\alpha)}+ 2^{N-2} s_2^{(\alpha)}+ \cdots + s_N^{(\alpha)}$ in terms of binary variables $s_i^{(\alpha)} \in \{0,1\}$. We define the tensor 
$F_{s_1 \ldots s_N }$ as,
\begin{align}
F_{s_1^{(\alpha)} s_2^{(\alpha)} \cdots s_N^{(\alpha)}} \equiv f_\alpha.
\end{align}
The last step is to parametrize the very large tensor $F_{s_1 \ldots s_N }$ in terms
of a MPS. Explicitly, it means that we introduce $2N$ matrices $M^i(s)$ with $i\in \{1,2\cdots,N\}$ and  $s \in \{0,1\}$ and write the tensor as,
\begin{align}
F_{s_1 \ldots s_N } =  
M^1(s_1) M^2(s_2)\cdots M^N(s_N).
\end{align}
The matrix $M^i(s)$ is of size $\chi_{i-1}\times \chi_i$ with by definition $\chi_0=\chi_{N+1}=1$. By taking $\chi\equiv\max_i \chi_i$ large enough (i.e. $\chi = 2^{N/2}$), any tensor $F$ can be put in the MPS form. The interest of the QTT representation stems from the fact that in many situations, a small value of $\chi$, independent of $N$, is sufficient to describe the function with high precision.

The generalization to $d$ dimensions can be done in several ways (see \cite{Jolly2025} for a discussion). Here we simply discretize the different variables, say $(x,y,z)$ in $d=3$ dimensions, separately into $2^{N/d}$ grid points per dimension. We obtain $d$ integers with $N/d$ bits each, e.g. $\alpha_x$, $\alpha_y$ and $\alpha_z$ in $d=3$ dimension. Each integer
is written in bit format and we form the tensor $F$ by putting the bits of the different dimensions one after the other.
In the example of $d=3$, we have:
\begin{align}
F_{s_1^{(\alpha_x)} \cdots s_{N/3}^{(\alpha_x)}
s_{1+N/3}^{(\alpha_y)} \cdots s_{2N/3}^{(\alpha_y)}
s_{1+2N/3}^{(\alpha_z)} \cdots s_{N}^{(\alpha_z)}
} \equiv f_{\alpha_x,\alpha_y,\alpha_z}.
\end{align}
Different orderings of the bits can be used (for instance one can group them scale by scale) but in the examples of this article we have not found that it affected the results very significantly.

Much is known about the QTT representation of various functions (e.g. cosine, polynomials...) as well as the MPO to
operate on a QTT (integration, differentiation, convolution, Fourier transform...) and we refer to the literature cited above for further references.

\begin{figure}[thb!]
  \centering
  \includegraphics[width=1.05\linewidth,trim = 0.0in 5.0in 2.2in 0.0in,clip=true]{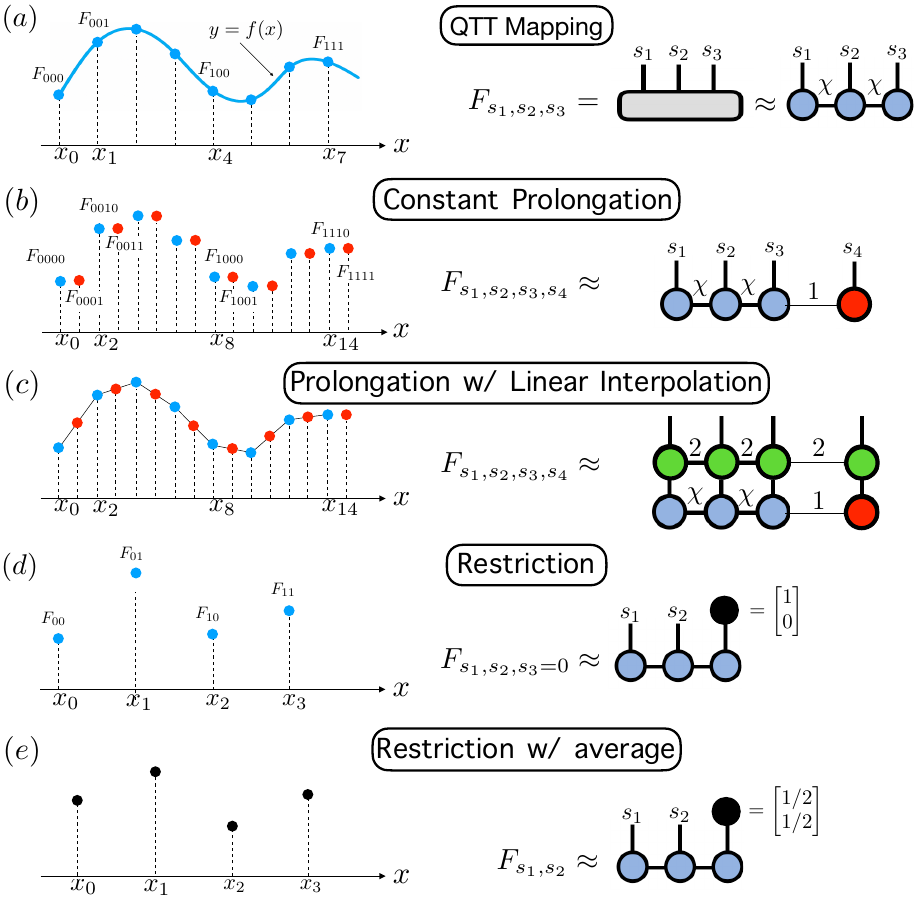}
  \caption{Schematic of the QTT representation and its change of resolutions. (a) The different objects in play: a univariate function, $y=f(x)$, its discretization onto $2^N$ points for $N=3$, the corresponding mapping to the tensor $F_{s_1s_2s_3}$ and finally the MPS representation of this tensor. (b) After the constant interpolation prolongation, a trivial bit $s_4$ is attached to the original QTT. (c) The linear interpolation is obtained by multiplying the elongated QTT from (b) with rank-$2$ interpolation MPO, see Appendix \ref{SI2}. 
  (d) A restriction move by projecting out the last bit on $0$. (e) Alternatively, one can choose to average over the last bit.  
  }
  \label{fig:I2}
\end{figure}

\subsection{Changing the resolution of a QTT}

To proceed, we need two important operations which we denote $\mathcal R$ (fine $\to$ coarse) and $\mathcal P$ (coarse $\to$ fine). The ``restriction''
$\mathcal R$ change $F$ with a $N$ bits resolution to a coarser QTT $F'=\mathcal R F$ with $N-d$ bits.
The ``prolongation'' $\mathcal P$ does the opposite and goes from $N$ bits to $N+d$ bits. Below we define these operators in $d=1$. The extension to higher dimension is straightforward. 
A cartoon for the two different restrictions and prolongations that we define is shown in Fig.\ref{fig:I2}.

\subsubsection{Restriction}
Applying $\mathcal R$ descends the resolution per dimension by one level from $N$ to $N-1$; it halves the spatial resolution by merging two neighboring fine grid points into a single one.
We consider two possibilities for the restriction. The first one is to simply ignore the value of $F$ on the points where $s_N=1$ and define $F' = \mathcal R F$ with
\begin{equation}
F'_{s_1s_2\cdots s_{N-1}} \equiv F_{s_1s_2\cdots s_{N-1}, 0}
\end{equation} 
or equivalently,
\begin{equation}
f'_\alpha \equiv f_{2\alpha}.
\end{equation} 
This restriction amounts to keep all the matrices $M^{\prime i}(s) = M^i(s)$ to be the same as those of $F$ except for
$M^{\prime N-1}(s) = M^{N-1}(s)M^N(0)$.

The second restriction amounts to averaging $F$ over the finer grid,
 \begin{equation}
F'_{s_1s_2\cdots s_{N-1}}  \equiv \frac{1}{2} \sum_s F_{s_1s_2\cdots s_N-1, s}.
\end{equation} 
or equivalently,
\begin{equation}
f'_\alpha \equiv \frac{1}{2} \left( f_{2\alpha} + f_{2\alpha+1}\right).
\end{equation}
This restriction amounts to keep all the matrices $M^{\prime i}(s) = M^i(s)$ to be the same as those of $F$ except for
$M^{\prime N-1}(s) = M^{N-1}(s)[M^N(0)+M^N(1)]/2$.

\subsubsection{Prolongation}
The prolongator $\mathcal P$, adds an additional binary bit to the QTT and doubles the number of grid points.
The simplest prolongation from $N$ to $N+1$ bits is to use a constant interpolation: We have $F'= \mathcal P F$ with,
\begin{equation}
F'_{s_1s_2\cdots s_N,0} = F'_{s_1s_2\cdots s_N,1} = F_{s_1s_2\cdots s_N}
\end{equation}
or equivalently
\begin{equation}
f'_{2\alpha} =  f'_{2\alpha+1} = f_{\alpha}.
\end{equation}
This prolongation amounts to keep all the matrices $M^{\prime i}(s) = M^i(s)$ to be the same as those of $F$.
The new matrix $M^{\prime N+1}(s)$ is a $1\times 1$ matrix defined as $M^{\prime N+1}(s) = 1$. This prolongation has the important drawback of introducing sharp discontinuities in both the first and second derivatives.

A better prolongation is to use a linear interpolation. In that case the points that were already present before prolongation are kept identical
$F'_{s_1s_2\cdots s_N,0} = F_{s_1s_2\cdots s_N}$, but the new point are obtained by linear interpolation between existing ones, $f'_{2\alpha+1} = (f_{\alpha}+f_{\alpha+1})/2$. A simple rank-2 MPO provides this linear interpolation so that the corresponding prolongation simply amounts to a standard MPO.MPS product. This MPO is a simple generalization of the MPO 
for the shift operator (see e.g. \cite{Waintal2026}). Higher order interpolants can also be constructed easily but we have found linear interpolation to be sufficient in the applications shown in the present article. Appendix \ref{SI2} shows the explicit construction of the prolongation for arbitrary high order interpolation schemes.

\subsection{Dynamical resolution solver}

\begin{figure}[!htb]
  \centering
  \includegraphics[width=1.0\linewidth,trim = 0.0in 2.0in 4.5in 0.0in,clip=true]{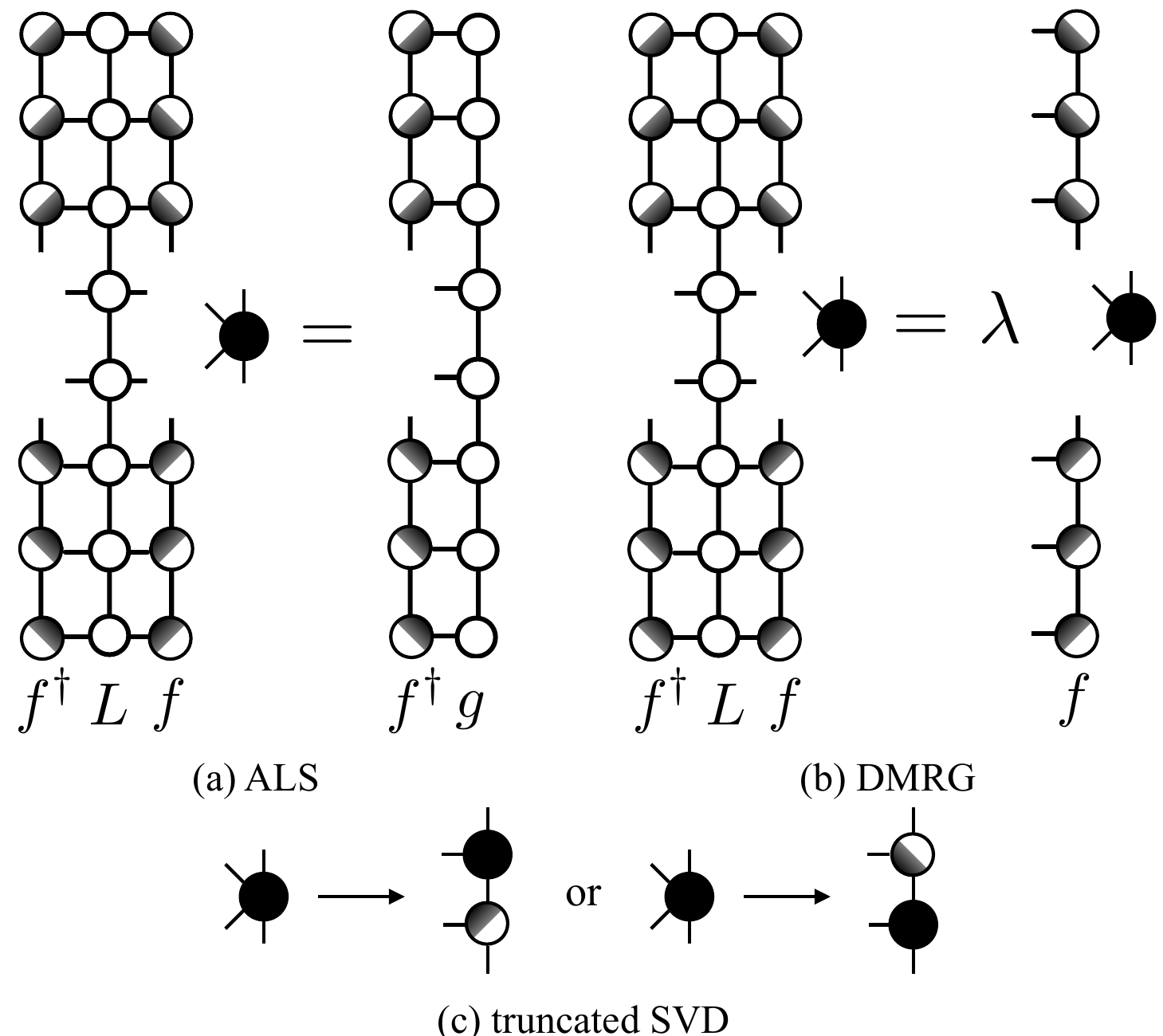}
  \caption{Schematics of the small local linear algebra problem that is solved during an ALS and DMRG calculation. A complete calculation consists of sweeping over the position of the $4$-legged black tensor until convergence. 
  	(a) Alternating least squares (ALS): the $4$-legged local tensor (in black) is singled out and optimized to minimize the cost function $|| f^{\dagger} L f - f^{\dagger} g ||$ while keeping the remaining tensors fixed.
  	(b) Density matrix renormalization group (DMRG) is similar to (a), but an eigenvalue problem is solved instead of a linear equation.
  	(c) Canonical form and truncated SVD: the MPS is brought into a mixed canonical gauge with \textit{orthonormal left} and \textit{right} blocks; the $4$-legged tensor is factorized via singular value decomposition, and small singular values are discarded to obtain an optimal low-rank approximation consistent with the canonical normalization conditions during left-to-right and right-to-left sweeps.  	
  }
  \label{fig:alg}
\end{figure}

The dynamical resolution solver combines the QTT and the multigrid approach in a straightforward way. Below we describe the algorithm to solve the Poisson
equation $\Delta f = \rho$ which, after discretization, maps to solving linear problems of the form $Lf = g$. The other example studied, solving a Schr\"odinger equation which maps onto eigenvalue problems $Lf = \lambda f$, follows exactly the same procedure except for the inner solver (ALS is replaced with DMRG). The different steps of the algorithm are:

\emph{(1) Initialization.} The input of the problem, the charge density $\rho (\mathbf{r})$
is mapped onto an MPS with the highest resolution $N_{\max}$ using a generalization of the Tensor Cross Interpolation (TCI) algorithm \cite{Oseledets2010, Oseledets2011, Savostyanov2014,Tindall2024}       as implemented in the xfac library \cite{NunezFernandez2025}. Applying the restriction $\mathcal{R}$ iteratively on the MPS of $\rho$ provides the MPS of $\rho$
at the coarsest level $N_{\min}$. In the case of the Poisson equation, it is interesting to use the restriction "with average" since, by construction, this restriction conserves the charge \emph{exactly}. Small variations of charge due to the discretization are often the main source of finite size errors in electrostatic simulations, so using the proper restriction can have a significant impact \cite{Armagnat2019}. This makes the method effectively a finite volume method. 

\emph{(2) Coarse solution.} Next, we build the linear operator $L$ at the coarsest level $N_{\min}$.
We choose a simple $2d+1$ points stencil for which the MPO of $L$ is known analytically and is of very low rank \cite{Waintal2026}. There is no need to apply the prolongation 
$L \rightarrow \mathcal{R} L \mathcal{R}$ here, although this could be done too if necessary. 
It is also straightforward to use a higher level approximation of the Laplacian but we have found it did not lead to obvious advantages (see the discussion later). The problem $Lf = g$ is solved at level $N_{\min}$ with a standard $2$-sites ALS solver (see \cite{Holtz2012,Oseledets2012b, White1992, Ostlund1995} for a detailed description) and a random initialization for the solution. 
The ALS solver is an iterative solver which contains the following steps: 
(i) one chooses a pair of consecutive tensors in the MPS of $f$. 
(ii) the MPS is put in the canonical form with the canonical center corresponding to pair of interest. 
(iii) one forms and solves a reduced linear problem as shown in Fig.~\ref{fig:alg}a . This is a small problem for which a standard linear solver (either a direct or a Krylov solver) can be used. 
(iv) the obtained tensor is separated using a singular value decomposition (SVD), see Fig.~\ref{fig:alg}c . 
(v) one sweeps the chosen pair by going up (and then down), repeating the steps (ii)-(iv) until convergence of the residual error $\| Lf-g\|\le \epsilon$ below a predetermined error level $\epsilon$.

\emph{(3) Interpolate and solve.} One uses the linear prolongation to obtain a guess of the solution of the problem at the next finer level. One uses the ALS solver to solve this problem.

\emph{(4) Repeat.} One repeats the step (3) until the resolution $N_{\max}$ has been reached.

The eigenvalue problem follows the same steps with two modifications: in step (iii) one
needs to solve the reduced eigenproblem of Fig.~\ref{fig:alg}b instead of the linear problem of Fig.~\ref{fig:alg}a, this is done using the Density Matrix Renormalization Group \cite{White1992, Ostlund1995}; in order to create the coarse version of $L$, one needs to apply the restriction to the potential too.


\section{Application to Poisson equation}
\label{sec:poisson}

\begin{figure*}[!htb]
  \centering
  \includegraphics[width=0.9\textwidth,trim = 0.0in 0.0in 0.0in 0.0in,clip=true]{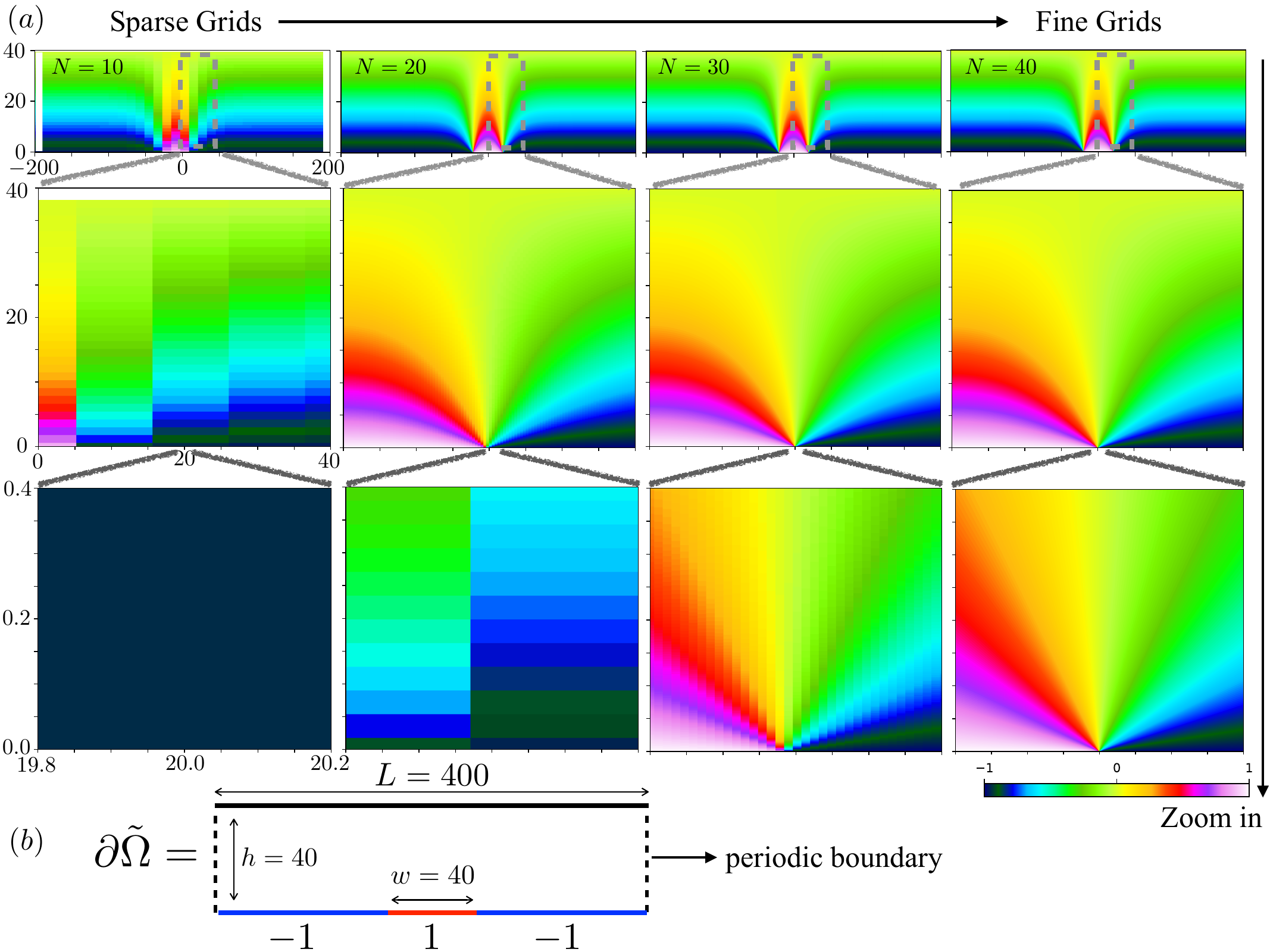}
  \caption{(a) Plots of the QTT solutions and its zoom-ins close to the discontinuity for Poisson's equation in two dimensions subject to the boundary condition depicted in (b) from coarse ($2^5 \times 2^5$) to fine resolution ($2^{20} \times 2^{20}$). (b) The boundary conditions: along the $x$ direction, the open boundary condition with discontinuities at the bottom. Along the $y$ direction, the periodic boundary condition is imposed. }
  \label{fig:F1}
\end{figure*}

\subsection{A 2D Benchmark}

We start with the two-dimensional problem shown in Fig.~\ref{fig:F1}b: a Poisson equation
in the absence of source $\Delta f(x,y) = 0$ with Dirichlet boundary conditions, 
\begin{itemize}
\item $\forall x,\ \ f(x,h)=0$,
\item for $|x|>w/2$, one sets $f(x,0)=-1$ and
\item for $|x|<w/2$, one sets $f(x,0)=+1$.
\end{itemize}
Since there is no charge density, the boundary conditions form the right hand side of the linear problem which is therefore very sparse.
Although non-trivial, this problem can be solved analytically using a conformal transformation and constitutes a good benchmark (see Appendix and \cite{Davies1995}). In particular its solution possesses sharp features around $\mathbf{x}=(-w/2,0)$ and $(+w/2,0)$ where the discontinuity of the boundary condition happens.

\begin{figure}[!htb]
  \centering
  \includegraphics[width=1.0\linewidth,trim = 0.0in 0.0in 0.0in 0.0in,clip=true]{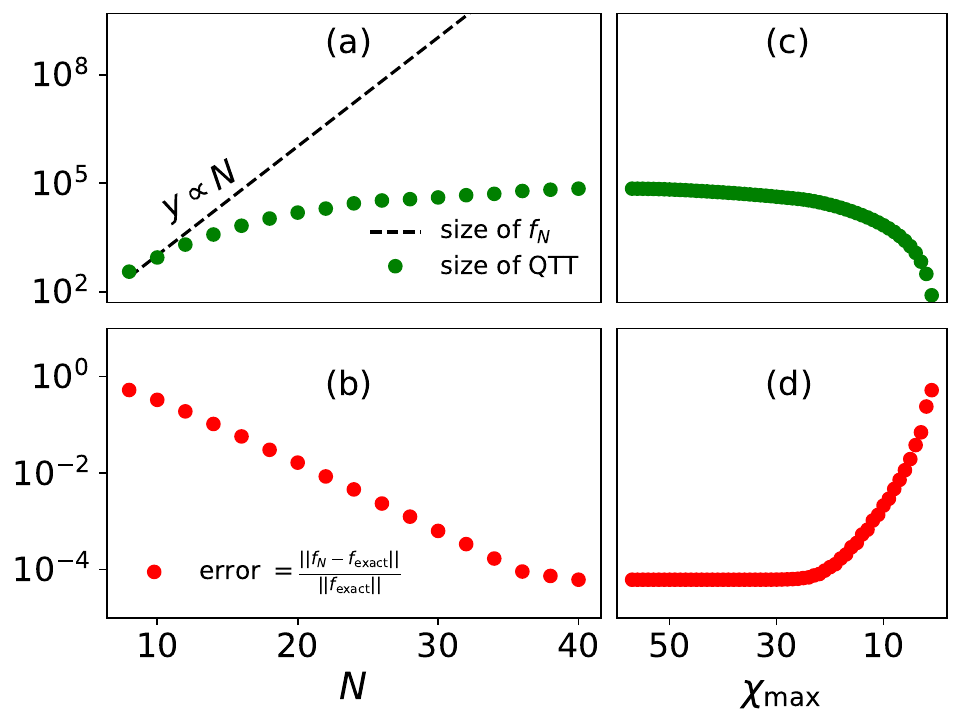}
  \caption{Complexity analysis of the Poisson's problem shown in Fig.~\ref{fig:F1}.
  \textbf{(a)} Number of parameters of the QTT solution (green circles) versus discretization level $N$. The dashed line shows $y=2^N$, the size without compression.
  \textbf{(b)} \textbf{red}: the relative error of the QTT solution versus $N$.  
  \textbf{(c) and (d)} Same as (a) and (b) respectively but versus the maximal bond dimension $\chi_{\max}$ at $N_{40}$. 
  }
  \label{fig:F2}
\end{figure}

Fig.~\ref{fig:F1}(a) shows our solution at different levels of resolution: from left to right $N=10$, $N=20$, $N=30$ and $N=40$. The first line shows the result of the entire simulation box while the second line shows a zoom close $\mathbf{x} = (+w/2,0)$ and the third line shows a zoom of the zoom. At the coarsest level, a $N=10$-site QTT captures the rough global feature, but $N=20$ is needed to observe the details above the central gate. (second row). Very close to the boundary between the central gate and the right gate (third row), we need the $N=30$ resolution level to obtain an approximate correct (but pixelated) solution and $N=40$ for this solution to be smooth. This $N=40$ solution represents a fine mesh grid of $2^{20} \times 2^{20} \sim 10^{12}$ grid points while the corresponding QTT contains only 
$71,080$ parameters to encode it. This amounts to a very high compression ratio $\sim 10^7$ while the error against the exact solution is well below $10^{-4}$ [see Fig.~\ref{fig:F2}(b)].

We study quantitatively the behavior of the Poisson solver in Fig.~\ref{fig:F2} by computing the number of parameters used (upper panels) and residual error (lower panels)
versus resolution $N$ (left panels) and maximum bond dimensions $\chi_{\max}$ (right panels). The error is defined as the $L_2$ norm of the difference between the exact solution and the obtained one.  This is a natural error for our problem since its calculation simply amounts to a scalar product between two MPS. 
This benchmark is particularly well suited for the QTT approach: the error decreases exponentially with $N$ while the number of parameters increases only linearly (a very mild bond dimension $\chi < 30$ is sufficient to reach the precision of $10^{-4}$).
The number of parameters required by the QTT representation is bounded by
$2\chi_{\max}^2 N $ which is very significantly smaller than $2^N$, the number needed without compression.

\subsection{Rapidly oscillating charge density}

A recurring question in the emerging field of QTT is to find use-cases where the QTT compression cannot be matched by other techniques. Consider, for instance, 
very smooth functions; they can be approximated by a polynomial with a precision that improves exponentially with the degree of the polynomial \cite{Trefethen2019}. For that class of functions, the high compression obtained with QTT simply reveals this smooth structure \cite{Lindsey2023}. In cases where only a few singular points are present, as in our above benchmark, one could argue that a combination of an adaptive grid (to refine the grid close to these points) with some basis of orthogonal (say Chebychev) polynomials could also solve the problem with very high resolution, hence that QTT is not really needed. These are valid points although we would be tempted to say that, even for these cases, the fact that the QTT representation is completely agnostic of the structure of the problem is in itself an important strength.

\begin{figure}[!htb]
  \centering
  \includegraphics[width=1.0\linewidth,trim = 0.0in 0.0in 6.0in 0.0in,clip=true]{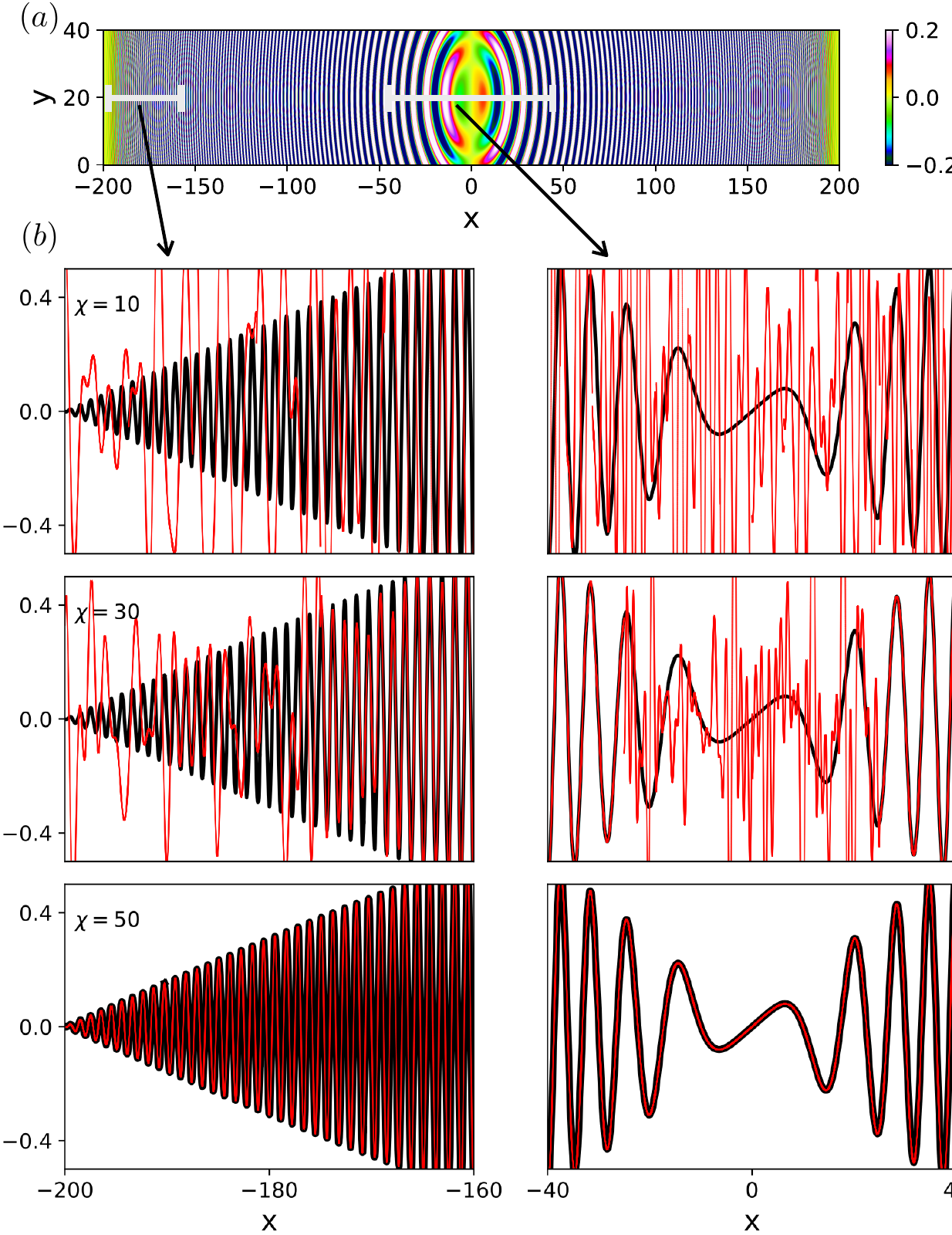}
  \caption{
  \textbf{(a)} Colormap of the oscillating charge density Eq.\eqref{eq:nasty_dst}.
  \textbf{(b)} $\rho(x,y)$ defined in Eq.\eqref{eq:nasty_dst} (black) and its QTT approximation (red) for different bond dimensions $\chi=10, 30$ and $50$ (top to bottom) and two 1D cuts of the 2D simulation box (left and right).}
  \label{fig:G1}
\end{figure}

Anyhow, in this section we consider an example that, to the best of our knowledge, could 
\emph{not} be addressed by these techniques. Although the problem that follows is somewhat contrived, it is inspired by homogenization problems where a media has rapidly varying properties at the microscopic scale but on is interested in solving the problem at a larger scale. Interestingly, often the theory that emerges at the larger scale has effective parameters that are highly non-trivial and they're \emph{not} simple averages of the microscopic ones. Any grid-based method must discretize the system with a grid finer than the smallest length scale and we therefore expect QTT to shine in this regime. So, we consider the following rapidly oscillating, yet non periodic, charge density
\begin{equation}
\label{eq:nasty_dst}
\rho(x,y) = \cos \left( 8\pi \tfrac{x^2+(y-h/2)^2}{h^2}\right) \times {\rm{sin}} (\pi \tfrac{x}{w} )
\end{equation}
A colorplot of this particularly nasty density is shown in Fig.~\ref{fig:G1}(a), it oscillates with a length scale comparable to the simulation box size near the center but much faster near the edges $|x|\sim 200$.

Eq.\eqref{eq:nasty_dst} is mapped onto its QTT form using TCI. Fig.~\ref{fig:G1}(b) shows the result given by TCI for three different values of the bond dimension $\chi= 10$ (upper row), $30$ (middle row) and $50$ (lower row) for the two different 1D cuts shown in panel (a) (red lines) respectively on the left (the charge density oscillates very rapidly) and in the center (mild oscillation). We observe that bond dimension $\chi=50$ is needed to account for the rapidly 
oscillating terms on the left. Overall, despite this function being relatively complex, it can be cast precisely in MPS form at the cost of a moderate bond dimension. 

\begin{figure*}[!htb]
  \centering
  \includegraphics[width=0.9\textwidth,trim = 0.0in 2.1in 0.0in 0.0in,clip=true]{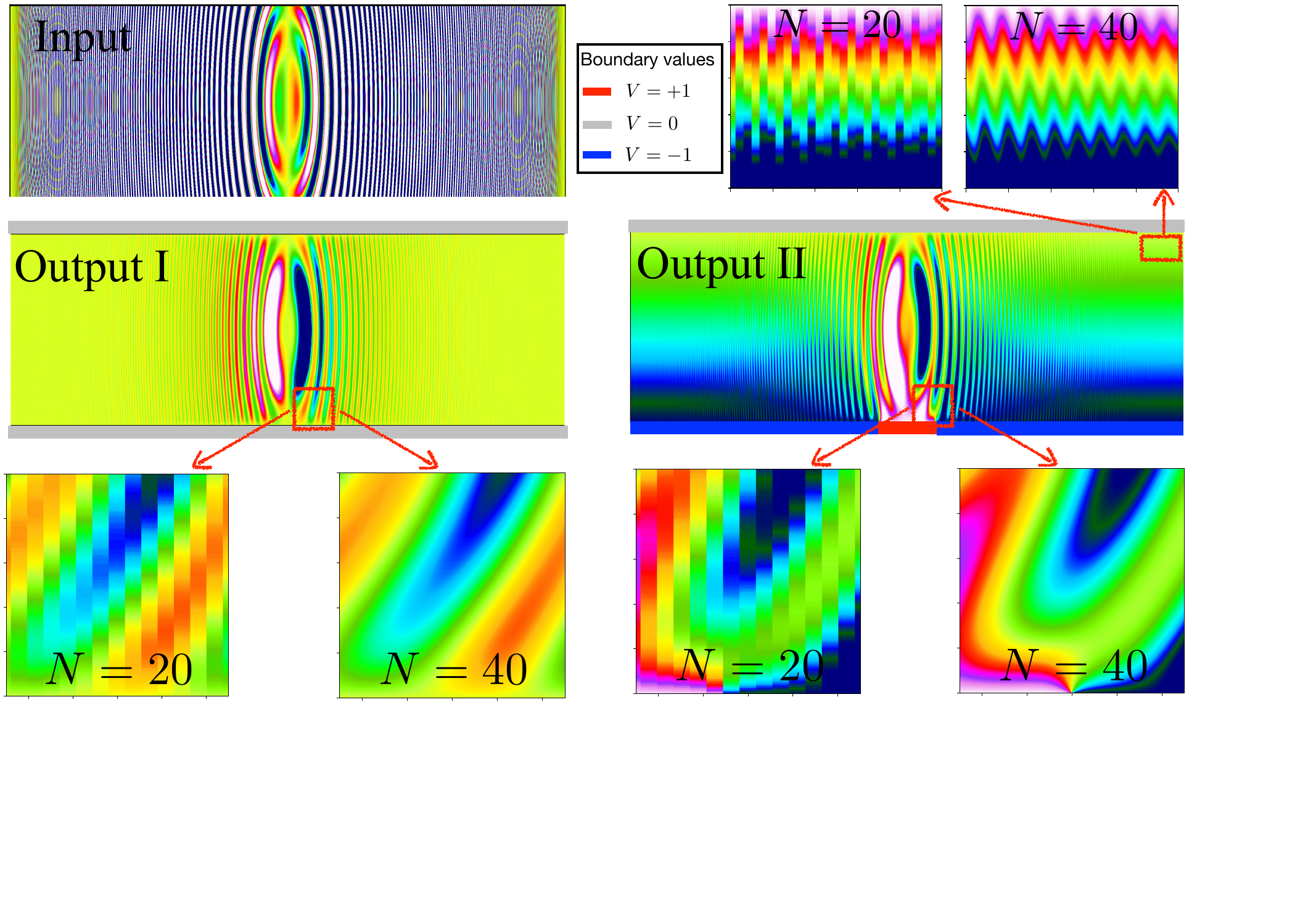}
  \caption{Two solution of Poisson's equation under different boundary conditions, with an input density $\rho(x,y)$ shown in Fig.~\ref{fig:G1}.
 }
  \label{fig:H2}
\end{figure*}

The solution of the resulting Poisson problem is shown in Fig.~\ref{fig:H2} with for two
different boundary conditions: $V=0$ everywhere for $y=0$ and $y=h$ (left panels) and
those of the above 2D benchmark (right panels). The zooms show the results at two different resolution $N=20$ and $N=40$ obtained at different stages of the resolution. As expected, the finer resolution is needed to properly grasp the fine details of the problem.

\begin{figure}[!htb]
  \centering
  \includegraphics[width=1.0\linewidth,trim = 0.0in 0.0in 0.0in 0.0in,clip=true]{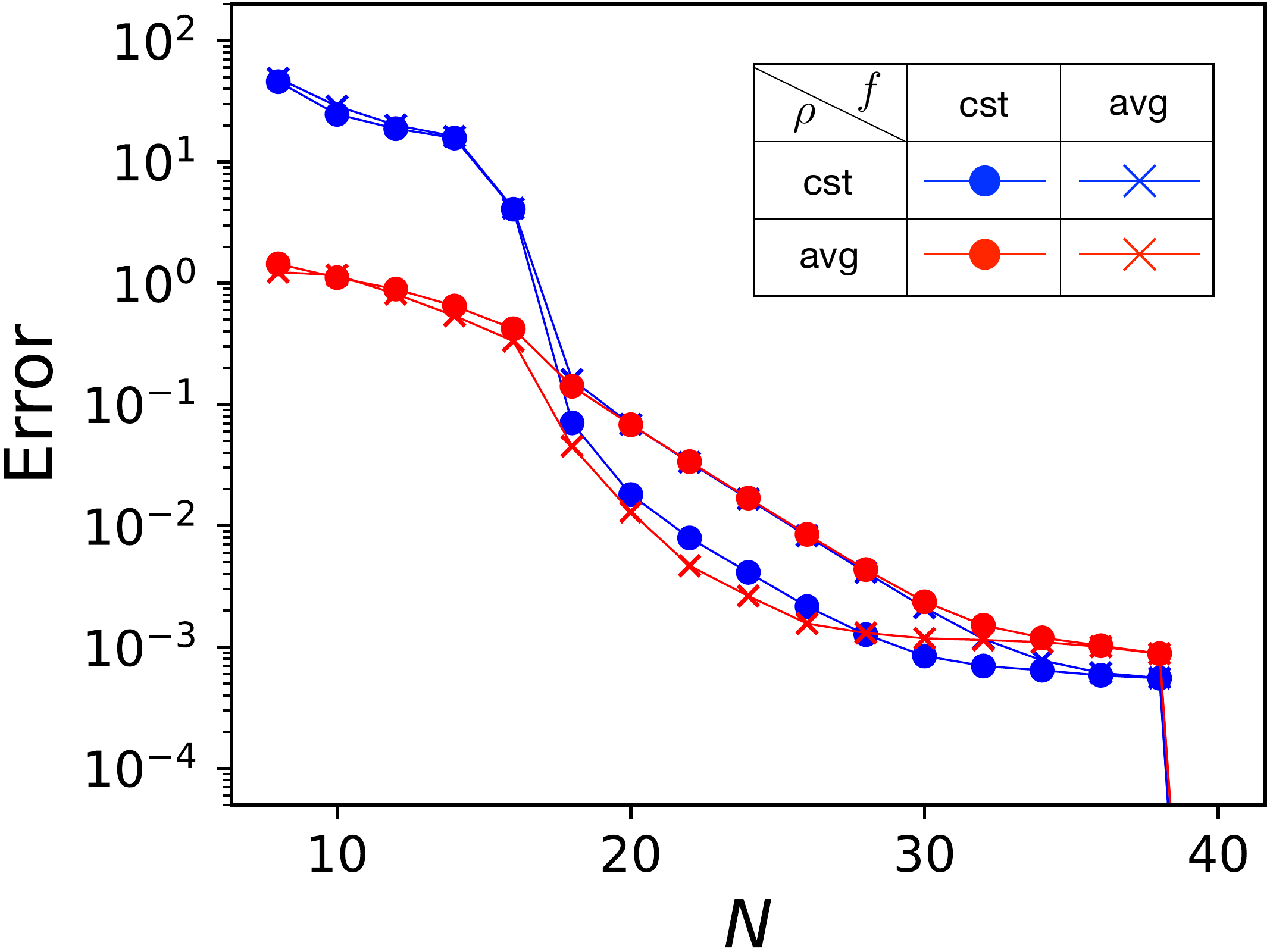}
  \caption{
  Error of QTT simulations for the 2D Poisson equation versus the number of total number bits, $N$ in the discretisation. At a given level $N$, we solve $\Delta f = \mathcal{R} \rho_{N=40}$ with two possible restrictions 
  $\mathcal{R} =\mathcal{R}_\text{cst}$ (constant restriction) and 
  $\mathcal{R} = \mathcal{R}_\text{avg}$ (averaging restriction).
 Similarly, we use two ways to compute the error,
 $\epsilon = || f(N)-\mathcal{R'} f_{N=40} ||/ ||f_{N=40}||$. with the same two possible 
 $\mathcal{R'}$.
  }
  \label{fig:errP}
\end{figure}

For this problem, we do not have a reference calculation, so we compute the error by comparing the solution at a given resolution with our converged solution $f_{N=40}$ at the final resolution $N=40$. More precisely, we perform two different calculations solving
\begin{equation}
\Delta f = \mathcal{R} \rho_{N=40}
\end{equation}
with $\mathcal{R} =\mathcal{R}_\text{cst}$ and $\mathcal{R} = \mathcal{R}_\text{avg}$ using respectively the constant and averaging restrictions. At a given scale, we calculate two types of errors 
\begin{equation}
\epsilon = || f(N)-\mathcal{R'} f_{N=40} ||/ ||f_{N=40}||,
\end{equation}
using constant (resp. averaging) restriction for $\mathcal{R'}$
 This gives us four combinations
of errors versus $N$ which are shown in Fig.~\ref{fig:errP}. Several points can be extracted from this figure. First, the two calculations converge to the same results within a precision of $\sim 10^{-3}$ which is a non-trivial consistency check. Second, the initial convergence of the error is faster with the run $\mathcal{R}_\text{avg}$: indeed within $\mathcal{R}_\text{cst}$, the coarse grid makes large finite size errors on the total charge inside the system which is highly problematic for electrostatic. Here the error is mitigated by the fact that the charge averages to zero but in a globally charged system, not accounting for all charges present could be catastrophic. Third, above a certain resolution $N\ge 18$, both runs show the same error when compared to their own converged result. The convergence of one run with respect to the final solution of the other one is slightly slower; it is a robust estimate of the true accuracy of the calculation.

\section{Application to a three body quantum problem}
\label{sec:schro}

We now turn to the three-body problem of calculating the spectrum of the ion $H_2^+$.
This is a benchmark problem that has been solved to high precision by using techniques that explicitly exploit its mathematical structure \cite{Sutcliffe1992, Cox1994, Ishikawa2008, Hijikata2009}. 
We will show that the quantics approach recovers these results with high precision while at the same time being able to deal with modifications that make the problem intractable with the above-mentioned techniques. 
All through this section, the DMRG calculations converged robustly with the multi-grid approach. We have found that the plain vanilla DMRG algorithm either did not converge at all as in Fig.~\ref{fig:Intro} or converged slowly requiring e.g. a slow increase of the bond dimension.

\subsection{Problem setup}

In its bare formulation, we seek to find the ground state and possibly some excited states 
of a system that consists of two nuclei situated at $\mathbf{r}_A$, $\mathbf{r}_B$ and
an electron at position $\mathbf{r}_e$. In atomic units, we want to solve the following 9D eigenvalue problem,
\begin{eqnarray}
&-&\frac{1}{2}\Delta_{\mathbf{r}_e} \bar\Psi
-\frac{1}{2m_A}\Delta_{\mathbf{r}_A} \bar\Psi
-\frac{1}{2m_B}\Delta_{\mathbf{r}_B} \bar\Psi
-\frac{Z_A}{|\mathbf{r}_e-\mathbf{r}_A|}\bar\Psi \nonumber \\
&-&\frac{Z_B}{|\mathbf{r}_e-\mathbf{r}_B|}\bar\Psi
+\frac{Z_AZ_B}{|\mathbf{r}_A-\mathbf{r}_B|}\bar\Psi
= E \ \bar\Psi(\mathbf{r}_e,\mathbf{r}_A,\mathbf{r}_B)
\end{eqnarray}
where $m_A$ (resp. $Z_A$) and $m_B$ (resp. $Z_B$) are the mass (resp. atomic number) of the two nuclei.
In the actual numerical experiments we restrict ourselves to two protons 
($Z_A=Z_B=1$, $m_A = m_B = 1836.152701$) but the method would stand for the more general problem. The above problems exclude the small relativistic and quantum electrodynamics
effect that can be accounted for using perturbation theory. Our first step is to rewrite the problem using the relative coordinates $\mathbf{R} = \mathbf{r}_A - \mathbf{r}_B$,
$\mathbf{r} = \mathbf{r}_e - [\mathbf{r}_A + \mathbf{r}_B]/2$ and the center of mass 
$\mathbf{C} = [\mathbf{r}_A + \mathbf{r}_B]/2$. Assuming that the center of mass is at rest, we obtain the reduced 6D problem,
\begin{eqnarray}
\label{eq:6D}
&-&\frac{1}{2}\left(1+\tfrac{1}{4\mu}\right) \Delta_{\mathbf{r}} \Psi
-\frac{1}{2\mu}\Delta_{\mathbf{R}} \Psi
-\frac{1}{2\mu_a}\mathbf{\nabla}_{\mathbf{r}}\cdot \mathbf{\nabla}_{\mathbf{R}} \Psi \\
&-&\frac{Z_A}{|\mathbf{r}+\mathbf{R}/2|}\Psi 
-\frac{Z_B}{|\mathbf{r}-\mathbf{R}/2|}\Psi 
+\frac{Z_AZ_B}{|\mathbf{R}|}\Psi
= E \ \Psi(\mathbf{r},\mathbf{R}) \nonumber
\end{eqnarray}
with
\begin{eqnarray}
\tfrac{1}{\mu} = \tfrac{1}{m_A} + \tfrac{1}{m_B} \\
\tfrac{1}{\mu_a} = \tfrac{1}{m_A} - \tfrac{1}{m_B} 
\end{eqnarray}
The second term vanishes in our numerical experiments with two identical nuclei.

In the following, we will solve this problem at three different levels of
approximation,
\begin{itemize}
\item \textbf{3D}. First, if one ignores the phonons of the problem in the limit 
$\mu\rightarrow\infty$ (Born approximation), one arrives at a 3D problem for a fixed value of $R=|\mathbf{R}|$. 
The resulting energy $E(R)$ can be optimized classically in a second step. This is the default approach used in quantum chemistry. It assumes that the electrons respond instantaneously to motion of the nuclei.
\item \textbf{3D+1D}. One can add the nuclei vibrations as an independent second step by solving the $1D$ model for the nuclei distance only,
\begin{equation}
\label{eq:1D_phonons}
-\frac{1}{2\mu}\Delta_{\mathbf{R}} \Psi + E(R)\Psi = E \Psi(R)
\end{equation}
Close to the classical minimum $R_{\min}$, $E(R)$ can be approximated by a parabola,
$E(R) \approx a (R-R_{\min})^2$. In that case, Eq.\eqref{eq:1D_phonons} reduces to a
harmonic oscillator of frequency $\omega \approx 2191.09952$ cm$^{-1}$. The resulting (non-variational) energy
is $E(R_{\min}) +\hbar\omega/2$ and the corresponding result is called \textbf{3D+HA}
(3D plus Harmonic Approximation).
\item \textbf{4D}. The full 6D problem is invariant with respect to a global rotation of both $\mathbf{r}$ and $\mathbf{R}$ so that there is a conservation of the global orbital momentum. Here we focus on the case where the state has no global orbital motion $l=0$,
which includes the ground state (and low energy excitations). The problem Eq.\eqref{eq:6D} reduces to a
4D problem where the vector $\mathbf{R}$ is replaced with a scalar $R$.  
\end{itemize}

\subsection{Representing the Coulomb kernel.}

\begin{figure}[!htb]
  \centering
  \includegraphics[width=1.0\linewidth,trim = 0.0in 0.0in 0.0in 0.0in,clip=true]{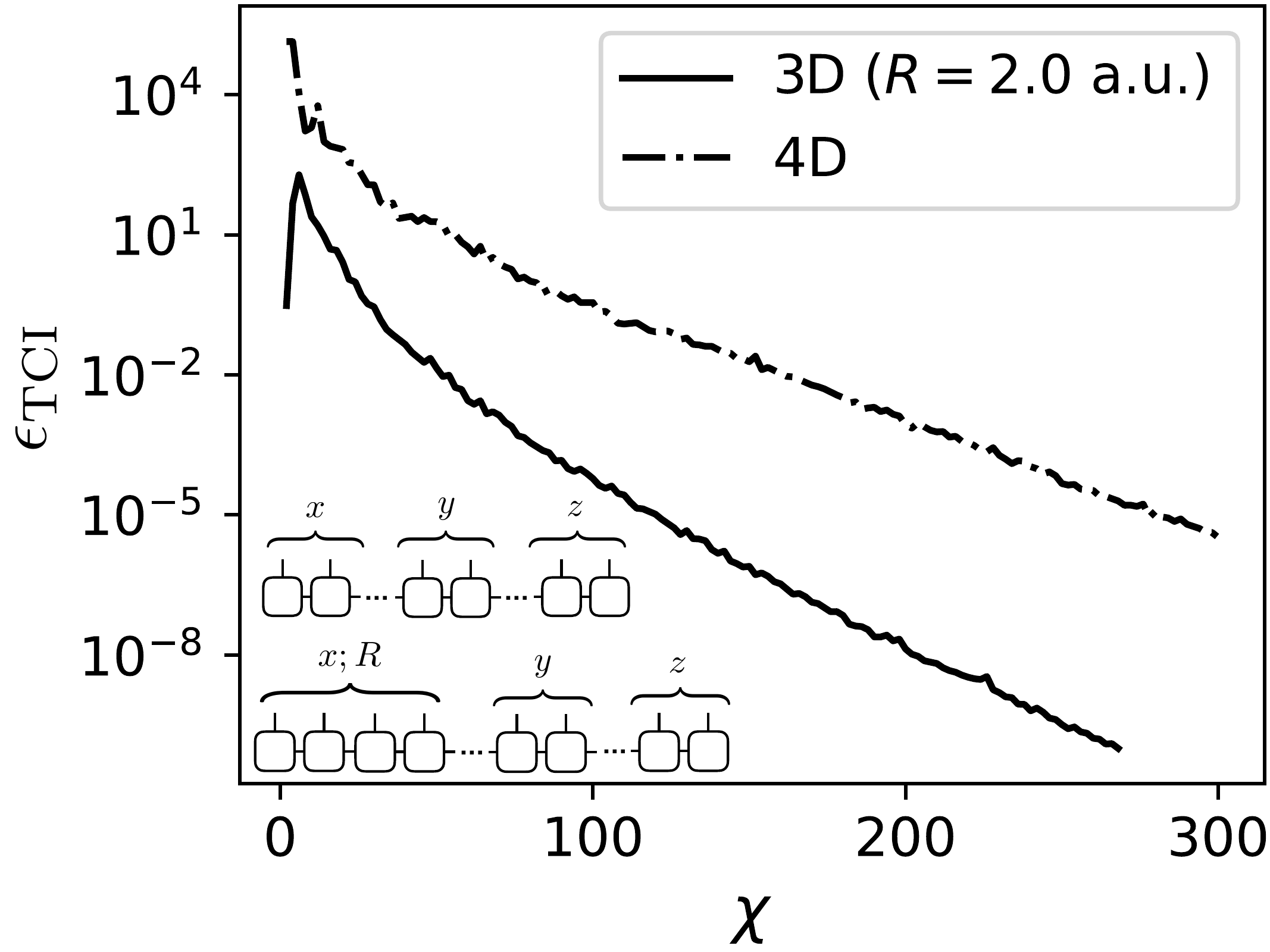}\
  \caption{ The TCI pivot error ($\epsilon_{\rm{TCI}}$) versus the maximal bond dimension ($\chi$) for approximating the Coulomb potential:  Eq.\eqref{eq:pot}.
  with $2^{20}$ uniform grid points for each variable.}
  \label{fig:H2+_tci}
\end{figure}

The kinetic part of the problem has a known exact expression in terms of a very small rank ($3$ in 1D) MPO \cite{Kazeev2012}. The difficulty here comes from the electrostatic potential $V(\mathbf{r},R)$ which diverges at small distance and decays slowly. Writing $\mathbf{r}=(x,y,z)$ we have,
\begin{equation}
\label{eq:pot}
 V(R,x,y,z) = \tfrac{1}{R} -\tfrac{1}{\sqrt{(x-R/2)^2+y^2+z^2}} - \tfrac{1}{\sqrt{(x+R/2)^2+y^2+z^2}} 
\end{equation}
To obtain the MPS representation of $V(\mathbf{r},R)$, we use TCI on the above function.
We have two different cases, see Fig.~\ref{fig:H2+_tci}.

For the 3D calculations, we fix the position of the nuclei, and TCI the potential considered as a function of $(x,y,z)$. In Fig.~\ref{fig:H2+_tci}, we fix $R=2$ and
use a grid of $2^{20}$ points for each of the three dimensions with 
$\mathbf{r} \in [-50,50]^3$. We use a sequential ordering of the bits to approximate the 
$60$-site tensor as a QTT,
\begin{equation}
V_{x_1x_2\ldots x_{20}\ y_1y_2\ldots y_{20}\ z_1z_2\ldots z_{20}}
\end{equation}
TCI converges relatively rapidly for the 3D problem, with an error
below $10^{-8}$ obtained with a bond dimension $\chi \sim 200$

For the 4D case, we TCI Eq.\eqref{eq:pot} considered now as a function of all four variables with $R\in [0.2,100.2]$. The small distance cutoff prevents the singularity while not affecting the results since the corresponding configuration has a prohibitively high energy.
We have found that a slightly different ordering of the bits is more favorable, so we "interleave" the $x$ and $R$ variables :
\begin{equation}
V_{R_1x_1R_2x_2\ldots R_{20}x_{20}\ y_1\ldots y_{20}\ z_1\ldots z_{20}}
\end{equation} 
The convergence of the QTT representation is slower than in 3D but nevertheless, as shown in Figure~\ref{fig:H2+_tci}, an error $\sim 10^{-6}$ is reached for $\chi=300$.

\subsection{Solving the 3D problem}

\begin{figure}[!htb]
  \centering
  \includegraphics[width=1.0\linewidth,trim = 0.1in 0.1in 0.1in 0.1in,clip=true]{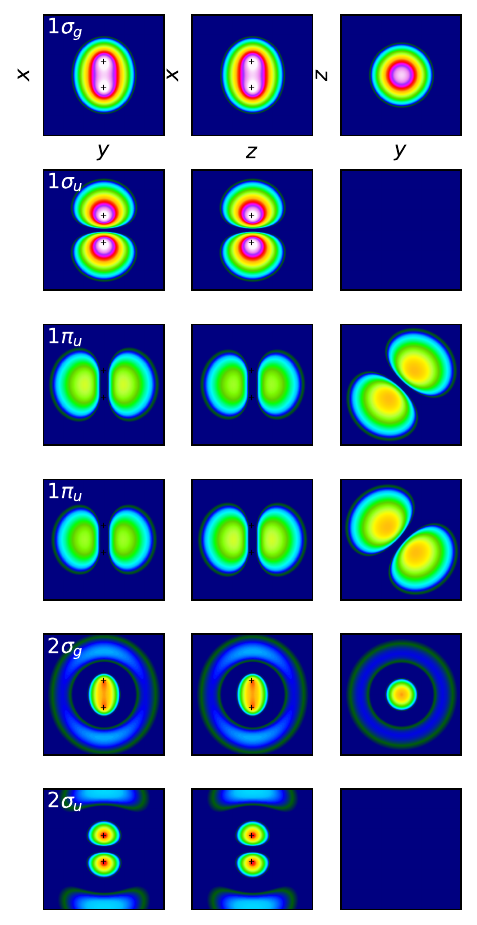}
  \caption{3D problem. two-dimensional density cuts of the electronic wavefunction $|\Psi(x,y,z)|^2$ along $x, y$ and $z$ directions for the first $6$ low-energy states for the $H_2^{+}$ with nuclei clamped at $(\pm1,0,0)$.}
  \label{fig:H2+3D_rho}
\end{figure}

The 3D problem has already been investigated by some of us using QTT using the default fixed grid approach \cite{Jolly2025}. We revisit this problem and take advantage of the improved convergence to calculate the low-lying excited states. To compute the $\alpha^{th}$ eigenstate $\Psi_\alpha$, we add a term $ +\mu\sum_{\beta<\alpha}\Psi_\beta\Psi_\beta^\dagger$ to the Hamiltonian. The Lagrange multiplier $\mu$ is chosen large enough to ensure the orthogonality between the previous eigenstates $\Psi_\beta$ and the targeted lowest-lying state. In Fig.~\ref{fig:H2+3D_rho}, we plot the electron density for the first six lowest energy eigenstates where the nuclei position is fixed at $(\pm1, 0,0)$.
It is worth noticing  that our calculations are performed in the Cartesian coordinate without prior knowledge of the system's symmetry. The orbital symmetry that arises from the prolate spheroidal coordinate is nevertheless well-recovered.
In fact, we can associate each eigenstate with its quantum numbers in agreement with the standard molecular theory. Here $\sigma$ (resp. $\pi$) is a shorthand for the orbital angular momentum of $0$ (resp. $1$) along the internuclear axis, and (un)gerade labels the (odd) even parity with respect to inversion symmetry.

\begin{figure}[!htb]
  \centering
  \includegraphics[width=1.0\linewidth,trim = 0.0in 1.2in 0.0in 0.0in,clip=true]{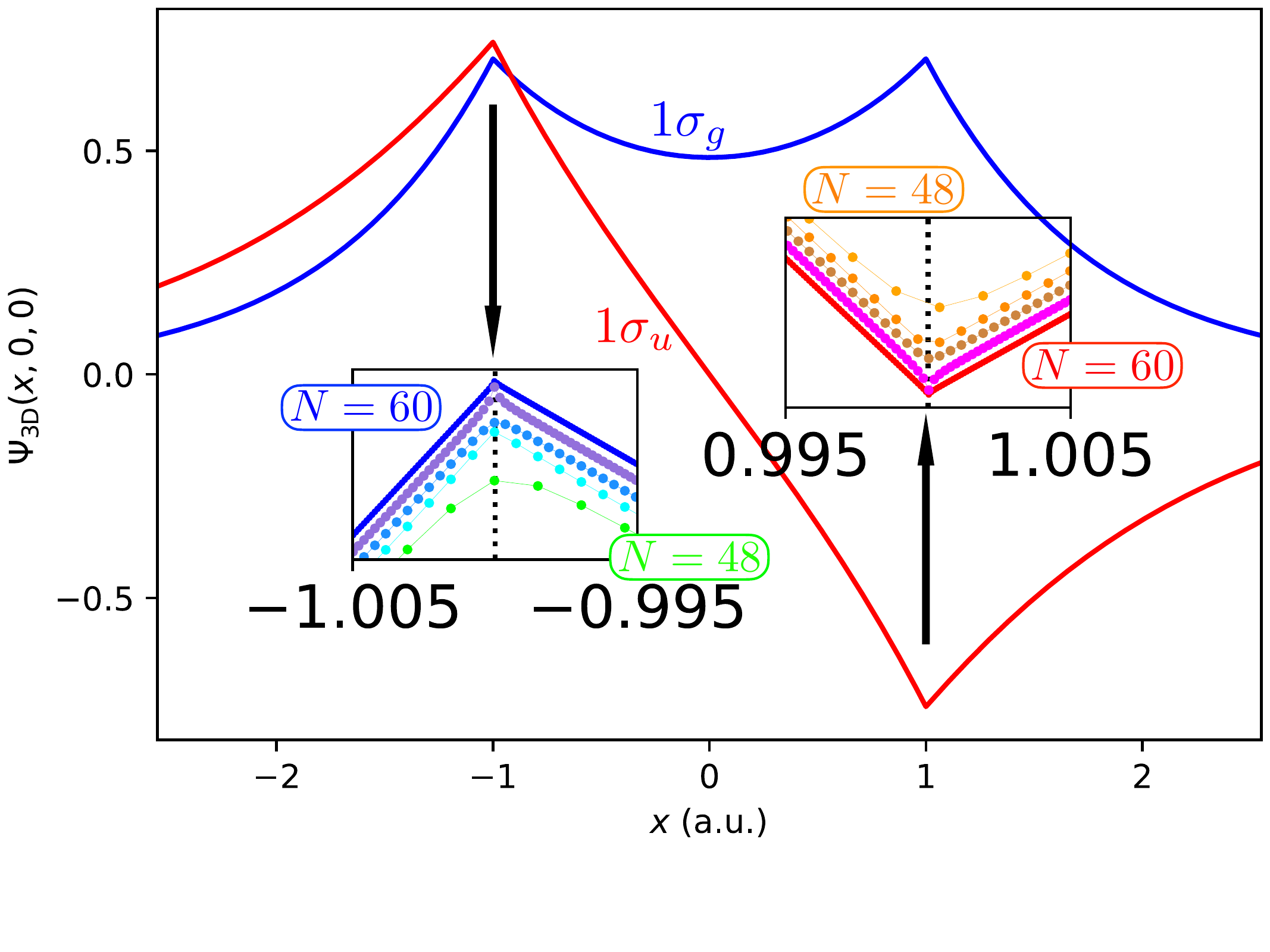}
  \caption{3D problem. One dimensional slice of the electronic ground state and the first excited state along the nuclear axis near the vicinity of singularities.
  Insets: zoom-in view to illustrate the cusp formation as the resolution is increased from $N= 48$ to $60$ (lines are off-shifted for better visualization).
  }
  \label{fig:H2+3D_wfn}
\end{figure}

Fig.~\ref{fig:H2+3D_wfn} plots 1D slices of the wavefunction along the nucleus axis $y=z=0$. The Coulomb potential implies that the wavefunction has cusps at the position of the nuclei
\cite{Kato1957,Bingel1967} which would be difficult to capture with, say, finite element methods. Our calculation showcase these cusps while the insets show clear deviations between $N=60$ and $N=48$ despite the latter already containing $> 10^{14}$ grid points. 

\begin{figure}[!htb]
  \centering
  \includegraphics[width=1.0\linewidth,trim = 0.0in 0.0in 0.0in 0.0in,clip=true]{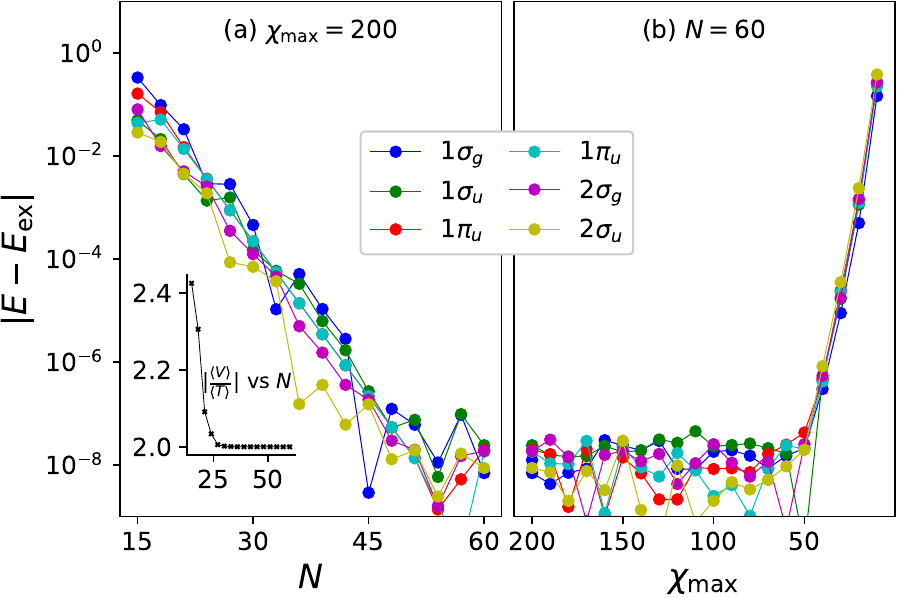}
  \caption{3D problem. Error on the energy (in unit of Hartree) for the first six low energy eigenstates (a) with respect to the total size $2^{N}$ of the grid with a fixed maximal bond dimension $\chi_{\rm{max}} = 200$ and (b) for the finite bond dimension effect $\chi_{\rm{max}}$ at $N=60$. Inset:  convergence with $N$ of the Virial ratio $\braket{V}/\braket{T}$  that relates the kinetic energy, $\braket{T}$, and potential energy, $\braket{V}$. For the ground state $\lim_{N\rightarrow\infty} \braket{V}/\braket{T}=2$ \cite{Fock1930}. 
  }
  \label{fig:H2+3D_ene}
\end{figure}

We benchmark the obtained eigenenergies against high-precision calculations, performed with generalized Hylleraas-Gaussian basis set and quadruple precision arithmetic \cite{Ishikawa2008}. Fig.~\ref{fig:H2+3D_ene} shows the energy error versus the discretization level $N$ (left panel) and maximum bond dimension 
$\chi_{\max}$ (right panel). For all eigenstates, we observe a rapid convergence of the energy until it reaches a maximum accuracy $\sim 10^{-7}-10^{-8}$. A very moderate bond dimension $\chi_{\max}\sim 50$ is sufficient to reach this accuracy but the resolution needs to be fairly high $N>50$. Since the QTT representation has a memory footprint $<2\chi_{\max}^2 N$, the compression ratio with respect to a naive representation is gigantic in this example. Here our ultimate precision is limited by our TCI representation of the Coulomb potential as well as numerical noise originating from the double precision arithmetic.

\begin{figure}[!htb]
	\centering
	\includegraphics[width=1.0\linewidth,trim = 0.0in 0.4in 0.0in 0.0in,clip=true]{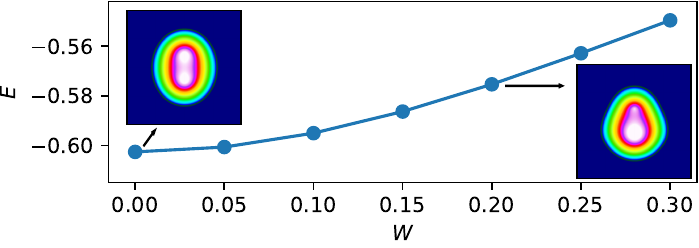}
	\includegraphics[width=1.0\linewidth,trim = 0.0in 0.0in 0.0in 0.0in,clip=true]{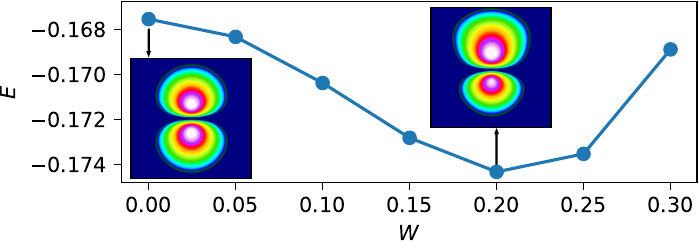}
	\caption{3D problem. Distortion of the energy of the ground state (upper panel) and the first excited state (lower panel) in the presence of an external oscillating electic potential of strength $W$.  The insets are snapshot of the wavefunction in the $z=0$ plane.}
	\label{fig:EXT}
\end{figure}

To conclude this section, we perform a calculation that breaks some of the symmetries of the problem: we place the di-hydrogen ion in an external electric field that oscillates along the $x$-axis, adding
\begin{equation}
V_{\rm{ext}}(\mathbf{r}) = W \sin\left(\dfrac{\pi}{2} x\right)
\end{equation}
to the Hamiltonian. This makes the eigenvalue problem more challenging to traditional techniques but does not affect our approach.
The results are shown in Fig.~\ref{fig:EXT} where we plot the energy $E(W)$ for the ground state (symmetric, upper panel) and the (anti-symmetric, lower panel) first excited state. We find that the energies move in the opposite direction when increasing the strength of the potential $W$ and that the method can capture the associated significant distortion of the orbital (see the corresponding snapshots). 

\subsection{Solving with the vibrations.}

\begin{figure}[!htb]
 	\centering
 	\includegraphics[width=0.9\linewidth,trim = 0.1in 0.1in 0.1in 0.1in,clip=true]{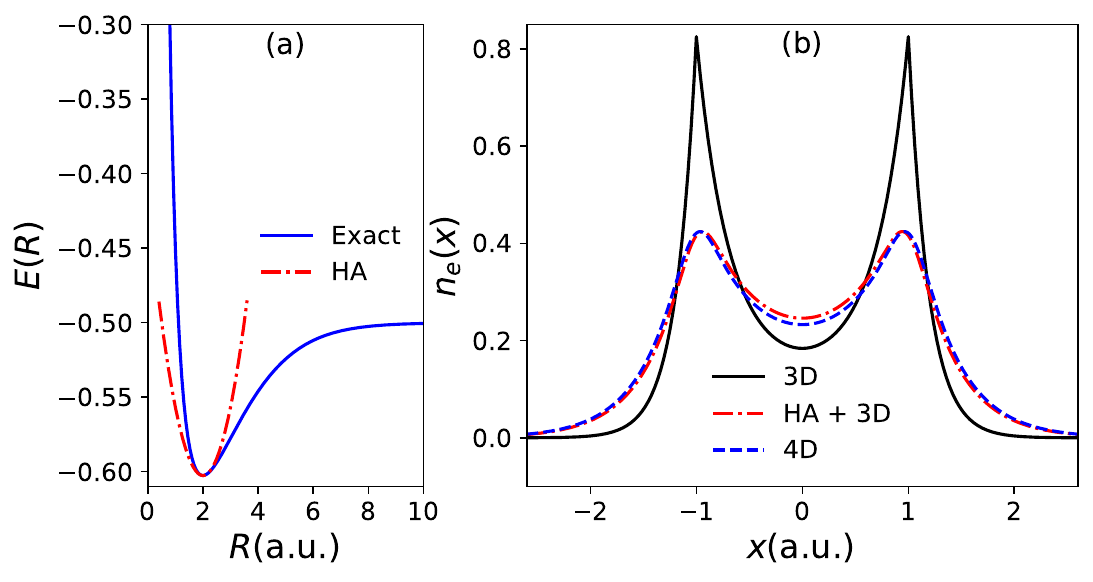}
 	\caption{(a) 3D+1D curve of electronic ground state energy versus $R$
 	(blue) and its HA fit (dashed red). (b) Electronic density $n_e(x)$ calculated at three different levels of approximations, see text.}
 	\label{fig:PH}
 \end{figure}

We continue with adding one new dimension and solve the 4D problem including the vibrations. For comparison and completeness, we also perform the 3D+1D calculation and the 3D+HA within our QTT framework (see Fig.\ref{fig:PH} for the obtained $E(R)$ 3D+1D energy curve together with its HA fit). 

We computed the
ground state energy $E_0$ as well as the first excited state energy $E_1$ in presence of a single phonon.
Table~\ref{tab:H2+} reports the deviation $\Delta E_{0/1}$ from the reference
values $E_0 = -0.597139$ Ha and $E_1 = -0.587155$ Ha calculated in the literature \cite{Hijikata2009}. The vibrational energy is of the order of 
$5$ mHa as shown by the value of the 3D problem that does not take it into account. $90\%$ of this energy is already accounted for by the 3D+HA approximation and $95\%$ by the 3D+1D calculation. Our $4D$ calculation reaches an additional order of magnitude in precision with an error of only  $0.02$ mHa for $N=80$ ($20$ bits per dimension)

\begin{table}[h]
\begin{center}
\caption{Accounting for the vibration energy of $H_2^{+}$ at different level of approximations. Deviations in energies (in unit of mHa) with 
respect to the literature values \cite{Hijikata2009}.}
\begin{tabular}{ ccccc}
	\hline
	\hline
	 & 3D & 3D+HA &  3D+1D &  4D \\ 
	\hline
	$\Delta E_0$ & $-5.4952$ & $-0.503$ &  $-0.256$ & $ +0.020$ \\ 
	$\Delta E_1$ & -- & $-0.504$ & $-0.253$ & $ +0.009$ \\
	\hline
\end{tabular}
\label{tab:H2+}
\end{center}
\end{table}

\begin{figure}[!htb]
	\centering
	\includegraphics[width=1.0\linewidth,trim = 0.0in 0.1in 0.00in 0.0 in,clip=true]{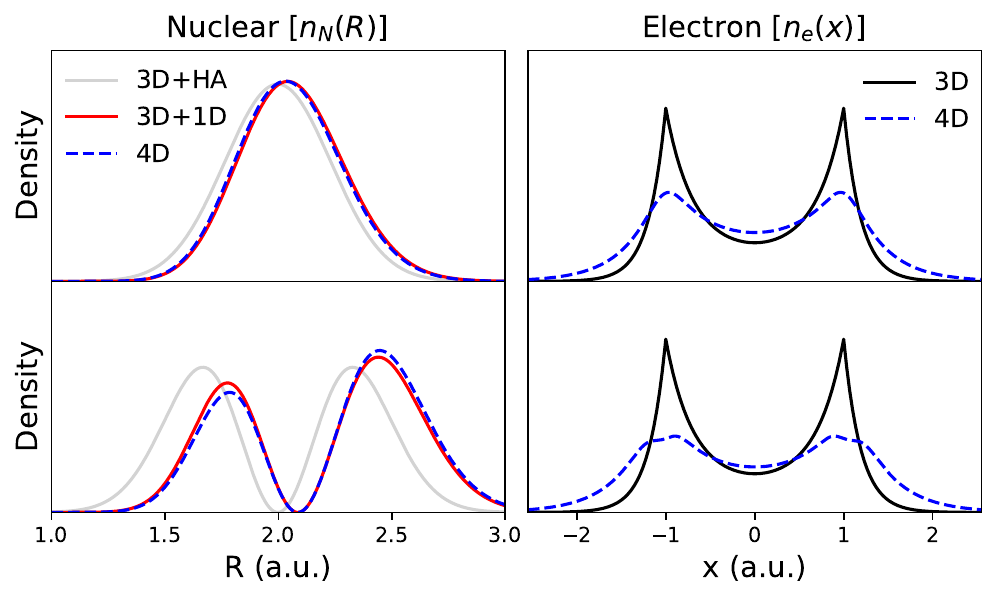}
	\caption{4D calculation of the $H_2^+$ ion. Left panels: nuclear density versus internuclear distance $R$. Right panels: electronic density at $\mathbf{r}=(x,0,0)$ versus position $x$. Upper panels: ground state. Lower panels: first vibronic excited state.
		The different curves correspond to the full 4D calculation (dashed blue), the 3D+1D calculation (red) and the 3D+HA approximation (gray). 
	}
	\label{fig:H2+_NR0}
\end{figure}

Being in possession of the full wavefunction $\Psi(\mathbf{r},R)$ for both the electron
and the nuclei position, we can now directly compute the electronic and nuclei density,
\begin{eqnarray}
n_e(\mathbf{r}) = \int dR |\Psi(\mathbf{r},R)|^2, \\
n_n(\mathbf{r}) = \int d\mathbf{r} |\Psi(\mathbf{r},R)|^2.
\end{eqnarray}
The results are shown in Fig.~\ref{fig:H2+_NR0} for the ground state (upper panels)
and first excited state (lower panels). The results for the nuclei density $n_n(\mathbf{r})$
(left panels) show that the harmonic approximation is already fairly accurate for the ground state but the system is already rather anharmonic as soon as 1 phonon is present in the system.
The 3D+1D calculation, however, is fairly accurate compared to the full 4D one.
The situation is drastically different when one looks at the electronic density
$n_e(\mathbf{r})$ (right panels). As expected, the cusp at the nuclei position present in the 3D calculation gets washed out when the quantum fluctuations of the nuclei are taken into account. However, the magnitude of the effect is quite important with a large redistribution of the electronic density close to the nuclei for both the ground state and the first excited state. The origin is clear: the zero point motion of the nuclei is very important in $H_2^+$ due to the lightness of the proton and weakness of the molecular link, so the electric potential seen by the electron is smeared out on that scale. What is also remarkable is how much the electronic density is affected by the presence of a single phonon
in the system (compare the upper right with the lower right panel): the redistribution of the electronic cloud is quite significant despite the fact that the energy only changed by
$\sim 10$ mHa.

The strong reorganisation of the electronic wavefunction due to zero point motion of the nuclei is a significant observation that may be harvested to improve both the accuracy
and the numerical stability of the 3D+1D calculation in an iterative scheme. Indeed, suppose we have solved the 1D problem and obtained $\Psi(R)$; the effective potential seen by the electron is
\begin{equation}
V_{\text{eff}}(\mathbf{r}) = \int dR \  |\Psi(R)|^2 \left(\frac{1}{|\mathbf{r}-\mathbf{R}/2|}  + \frac{1}{|\mathbf{r}+\mathbf{R}/2|}  \right).
\label{eq:V_eff}
\end{equation}
This potential has two advantages with respect to the bare Coulomb one: first, it is more accurate as it already includes the vibrations of the nuclei to a certain level;
second, it has smeared the short distance $1/|\mathbf{r}|$ divergence that is costly to account for numerically in our scheme. 
As a first step in the direction of performing a "self-consistent 3D+1D" calculation, we have solved the 3D problem where the Coulomb potential is replaced with Eq.~\eqref{eq:V_eff} assuming $\Psi(R)$ is described by the harmonic approximation. The result of this "HA+3D" calculation is 
shown in Fig.~\ref{fig:PH}b. For contrast, we also show the 3D and the 4D calculations. We find that this two steps approach, while still being  approximate, is actually very close to the full 4D result.
 
\section{Conclusion}

The QTT representation has already shown a number of interesting proofs of concept
as a powerful approach to solve partial differential equations. We are now entering the maturation stage where these techniques must go from the proof of concept stage to becoming a robust tool that one will use for practical use cases. Towards this goal,
adapting relevant ideas that have been successful in the context of other partial differential equation approaches is a rather natural route. In this work, we have explored the fact that the QTT representation has a natural multigrid interpretation to take a step
in that direction in the spirit of \cite{Lubasch2018}. We have found that the calculation could indeed be stabilized quite significantly as demonstrated by our 4D calculation where we obtain a meaningful accuracy of $\sim 10^{-4}$ for a system that involves $2^{80}\sim 10^{24}$ grid points, far beyond the capabilities of standard grid-based methods.
Future work will involve more advanced prolongations (e.g. higher order), more advanced
cycles, schemes where the MPS rank also evolves with the resolution, extension to a larger class of problems and more. It is our belief that QTT techniques are already above the threshold where one could boast of a "quantum inspired advantage".

\begin{acknowledgments}
Authors acknowledge the funding of Plan France 2030 ANR-
22-PETQ-0007 “EPIQ”, the PEPR “EQUBITFLY”,the
ANR “DADI”, the ANR TKONDO and the CEA-FZJ French-German project AIDAS for funding.
\end{acknowledgments}

\appendix

\section{Analytical solution of the benchmark Poisson problem}
\label{app1}

In this Appendix, we derive the analytical solution we used
in our Poisson benchmark. We follow a standard procedure
for solving 2D problems with conformal transformations.

\begin{figure}[!htb]
	\centering
	\includegraphics[width=1.0\linewidth,trim = 0.0in 7.9in 0.00in 2.0 in,clip=true]{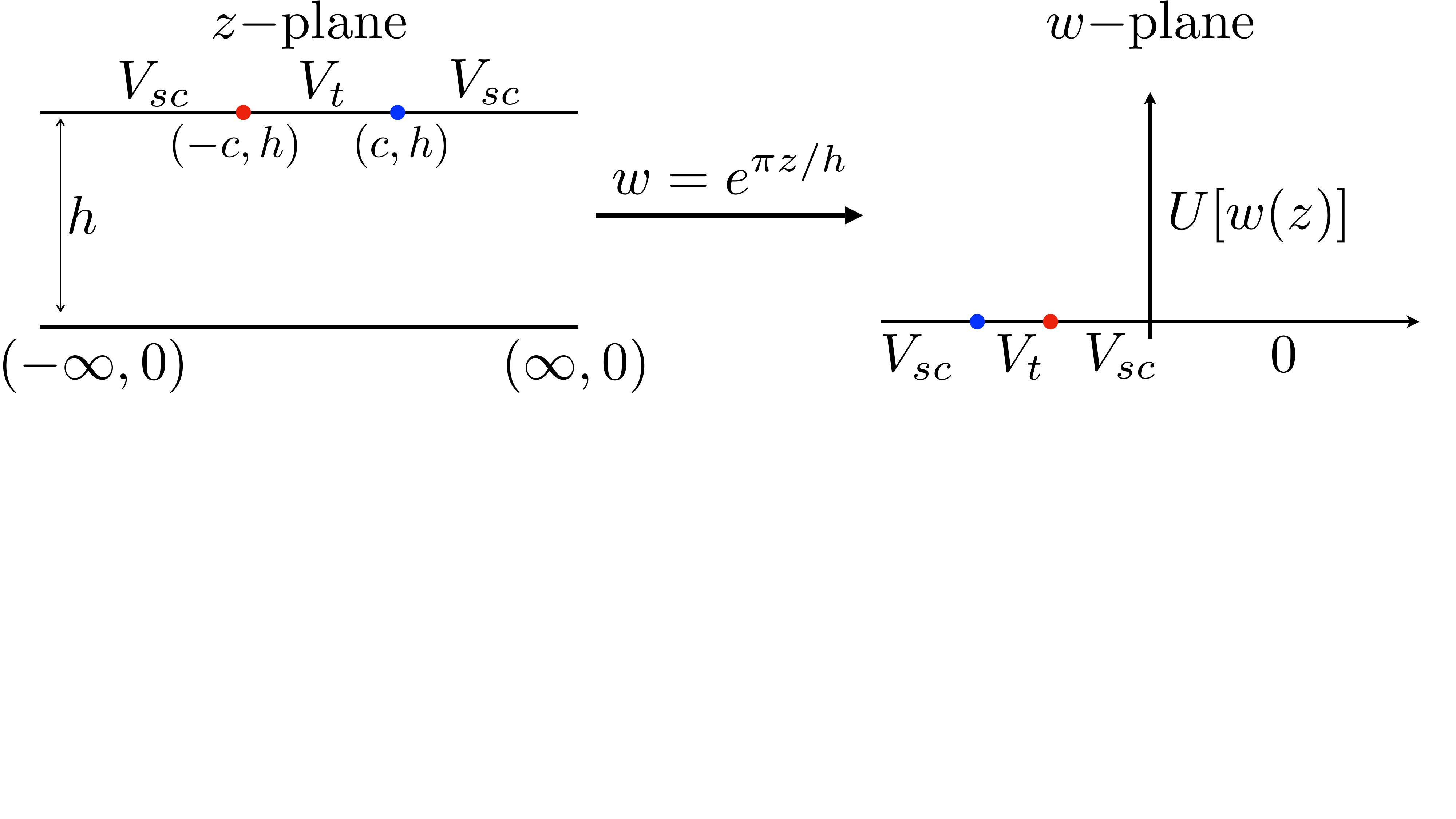}
	\caption{The conformal transformation used to map the strip geometry ($z$-plane) onto the upper half-plane ($w$-plane)}
	\label{fig:conformal}
\end{figure}

\subsection{Problem formulation}
We seek the solution of $\Delta V(x,y) = 0 $ inside a stripe
of width $h$ ($0<y<h$) infinite along the $x$ direction. We use
Dirichlet boundary conditions with
\begin{subequations}
\label{eq:appA1}
\begin{align}
V(x,0)&=0, \\ 
V(x,h)&=
\begin{cases}
V_t, & |x|<c \\
V_{\rm sc}, & |x|>c 
\end{cases}
\end{align}
\end{subequations}

\subsection{Conformal transformation}

Introducing $z=x+iy$, any analytical function $V(z)$ automatically satisfies
$\Delta V = 0$. We consider the exponential map $w(z) = e^{\pi z / h}$
that maps our stripe onto the upper half plane $w= u + iv$ with $v\ge 0$.
A schematic of the transformation is shown in Fig.~\ref{fig:conformal}.
We get the following mappings $z\rightarrow w$,
\begin{eqnarray}
(+\infty,0) &\rightarrow& (\infty,0) \nonumber \\
(-\infty,0) &\rightarrow& (0,0) \nonumber \\
(-\infty,h) &\rightarrow& (0,0)\nonumber \\
(-c,h) &\rightarrow& (-a,0)\nonumber \\
(c,h) &\rightarrow& (-b,0)\nonumber \\
(\infty,h) &\rightarrow& (-\infty,0)\nonumber 
\end{eqnarray}
with $a = -e^{\pi c/h}$ and $b = -e^{-\pi c/h}$.
We are left to solve the problem $\Delta U(u,v) = 0$ on the upper half plane with 
the boundary conditions,
\begin{align}
U(u,v=0)=
\begin{cases}
0, & u>0,\\
V_t, & u\in [a,b],\\
V_{\rm sc}, & u\in ]-\infty,a]\cup [b,0].
\end{cases}
\end{align}
Such a solution provides the solution to the original problem through $V(z) = U[w(z)]$.

\subsection{Solving the upper half plane problem}
Fortunately, the solution of the problem on the half plane is known for general boundary conditions. For $U(\xi,v=0)=G(\xi)$, it reads \cite{Axler2001}, 
\begin{equation}
U(u,v) = \frac{1}{\pi} \int_{-\infty}^{+\infty} d\xi \frac{v G(\xi)}{v^2 + (u-\xi)^2}.
\end{equation} 
In the particular case where $G(\xi <0)=1$ and $G(\xi \ge 0)=0$, it translates into
\begin{equation}
U(w) = \frac{1}{\pi} \text{Arg} (w) = \frac{1}{\pi} \arctan(v/u)
\end{equation}
as one can easily verifies explicitly. It follows that the solution to our problem in the
$w$-plane reads,
\begin{align}
U(w)= \frac{V_{\rm sc}}{\pi}{\rm{Arg}}(w) + \frac{V_t-V_{\rm sc}}{\pi}
[\text{Arg}(w-b) - \text{Arg}(w-a) ].
\end{align}
Replacing with the expression of $w$ in terms of $z$, we arrive at the solution
that matches the piecewise constant boundary conditions Eq.~\eqref{eq:appA1}.
\begin{align}
V(x,y) = \frac{V_{\rm sc}\,y}{h}
+ \frac{V_t-V_{\rm sc}}{\pi}\Big(\arctan (A_1)-\arctan (A_2)\Big).
\end{align} 
Here, we have defined, 
\begin{eqnarray}
A_1(x,y)=\frac{e^{\pi(c-x)/h}+\cos\!\big(\tfrac{\pi y}{h}\big)}{\sin\!\big(\tfrac{\pi y}{h}\big)},\\
A_2(x,y)=\frac{e^{-\pi(c+x)/h}+\cos\!\big(\tfrac{\pi y}{h}\big)}{\sin\!\big(\tfrac{\pi y}{h}\big)}.
\end{eqnarray}

\section{Explicit construction of the prolongation}
\label{SI2}

In this Appendix, we construct explicitly the averaging prolongation.
First, we do it in a pedestrian way and show that the averaging can be performed by a rank 2 MPO. Second, we use a slightly more abstract formulation that allows one to generalize the prolongation to arbitrary higher orders interpolations.

\subsection{A pedestrian construction}
We start from the constant prolongation where $f'_{2\alpha+1} = f'_{2\alpha}= f_{\alpha}$
and want to convert it into the linear average prolongation. This is done by performing the averaging $f'_{2\alpha}\rightarrow [f'_{2\alpha} + f'_{2\alpha+1}]/2$ on all points odd and even, the operation leaving the even points unchanged. 

We want to write the operator that performs this operation in the form of a rank $2$ MPO $S$ following closely what is done for the so-called shift operator.  
It takes the form
\begin{align}
\hat W = v_{\rm L}\,
\Big(\prod_{j=1}^N W^{[j]}\Big)\,
v_{\rm R},
\end{align}
where each element $W^{[j]}_{cc'}$ (the index $c,c'\in \{0,1\}^2$ we will be called the "carry" in what follows in analogy to how additions are taught to children) is itself a $2 \times 2$ matrix that acts on bit $j$. 
Introducing the $2 \times 2$ identity matrix $I$ as well as the Pauli matrices $\sigma_X$, $\sigma_Y$ and $\sigma_Z$, we define the raising  $S^+ = (\sigma_X+i\sigma_Y)/2$ and lowering 
$S^+ = (\sigma_X-i\sigma_Y)/2$ operators that respectively turn bit 0 into 1 and vice versa.
Before giving the general form of the matrix $W^{[j]}$, let us start with a few concrete examples to understand what the matrices should do. We only consider the shift part of the operator $(\hat W f)_\alpha = f_{\alpha+1}$ (the constant part is trivial). Suppose the input tensor is 
$$F_{s_1\cdots s_N} = \delta_{s_1,0} \delta_{s_2,0} \cdots \delta_{s_N,0}$$.
Then, we want  $(\hat W F)_{s_1\cdots s_N}$ to be equal to 
$(\hat W F)_{s_1\cdots s_N} = \delta_{s_1,0} \delta_{s_2,0} \cdots \delta_{s_N,1}$, i.e.
the MPO should act as $I\cdots I S^+$, i.e. raise the last bit and do nothing else.
If on the other hand we take a different input tensor
$$F_{s_1\cdots s_N} = \delta_{s_1,1} \delta_{s_2,1} \delta_{s_3,0} \delta_{s_4,1} \cdots \delta_{s_N,1}$$, 
then we want  $(\hat W F)_{s_1\cdots s_N}$ to be equal to 
$(\hat W F)_{s_1\cdots s_N} = \delta_{s_1,1} \delta_{s_2,1} \delta_{s_3,1} \delta_{s_4,0} \cdots \delta_{s_N,0}$ i.e. the MPO should act as $I I S^+ S^-\cdots S^-$. In short, starting from the least important bit, we want to lower all the bits equal to one until we find the first bit equal to zero (which is raised) and then we do nothing to the remaining bits. This is achieved by using a matrix $W^{[j]}$ that takes the form,
\begin{align}
W^{[j]} = \begin{pmatrix}
W_{0,0} & W_{0,1} \\[4pt]
W_{1,0} & W_{1,1}
\end{pmatrix}
=
\begin{pmatrix}
I & S^+ \\[4pt]
0 & S^-
\end{pmatrix}.
\end{align}
We further suppose that
\begin{align}
v_{\rm L} = \begin{pmatrix} \tfrac{1}{2} & 0 \end{pmatrix}, \qquad
v_{\rm R} = \begin{pmatrix} 1 \\ 1 \end{pmatrix}.
\end{align}
To understand why the above definition does indeed do what we need, let
us start from $v_{\rm R}$ and follow what happens when we multiply on the left
by $W^{[N]}$, then $W^{[N-1]}$,...i.e. starting from the least significant bit.
Suppose first that $c'=0$ ("no carry"). Then, by construction the tensor vanishes unless $c=c'$ so that $W^{[j]}_{cc'} = I$ for all bits. This is the constant part of the interpolation.
Suppose now that $c'=1$ ("active carry"). Then two things can happen:
if the bit $j$ is in state zero then $W^{[j]}_{01} = S^+$ will raise it (the carry goes back
to zero therefore all subsequent operators will be identities); 
if the bit $j$ is in state one then $W^{[j]}_{11} = S^-$ will lower it 
but we need to "carry" on to the next tensor, so $c=1$. 

\subsection{Generalization to higher orders}

The previous construction can be generalized to interpolations to arbitrary orders.
For instance, we would like to use a fourth order interpolation scheme like this,
\begin{eqnarray}
f'_{2\alpha} &=& f_{\alpha} \\
f'_{2\alpha+1} &=&  \tfrac{1}{6}\left( -f_{\alpha-1} + 4 f_{\alpha} 
+ 4 f_{\alpha+1} - f_{\alpha+2} \right) 
\end{eqnarray}
For this, we follow a slightly different route and do not start with performing the constant extrapolation. Instead, we proceed as follows:
\begin{itemize}
\item Construct an MPS for 
\begin{equation}
g_\alpha = \tfrac{1}{6}\left( -f_{\alpha-1} + 4 f_{\alpha} 
+ 4 f_{\alpha+1} - f_{\alpha+2} \right)
\end{equation}
using a multiple-shift MPO (the construction of which is explained below). We note
$G^i(s_i)$ the matrices of this MPS while, we recall, those of $f_\alpha$ are noted
$M^i(s_i)$.
\item Construct an MPS that contains both $f_\alpha$ and $g_\alpha$. The matrices of this MPS are given by,
\begin{equation}
\begin{pmatrix}
M^i(s_i) & 0 \\
0 & G^i(s_i)
\end{pmatrix}
\end{equation}
At this stage it is not exactly an MPS because the number of columns of the last matrix
\begin{equation}
\begin{pmatrix}
M^N(s_i) & 0 \\
0 & G^N(s_i)
\end{pmatrix}
\end{equation}
is equal to $2$ (one that addresses $f_\alpha$ and one for $g_\alpha$).
\item We complete the extrapolation with $2\times 1$ matrices $M^{N+1}(s_{N+1})$
defined as
\begin{equation}
M^{N+1}(0) =\begin{pmatrix}
1 \\
0 
\end{pmatrix},
\ \ \ 
M^{N+1}(1)=\begin{pmatrix}
0 \\
1 
\end{pmatrix}
\end{equation}
to select the right value depending on the parity of the added bit. We obtain the MPS of $f'_\alpha$.
\item We perform a standard compression of the MPS using SVD in the canonical basis to remove any spurious component. This ends the construction of the high order prolongation.
\end{itemize} 

We are left with the construction of the multiple-shift MPO. Its construction is done
using the "magic tensor" described in section 10 of \cite{Waintal2026}. This is a 5 indices tensor $M_{zxycc'}$ that automatically capture the constraints of adding two numbers whose
bits correspond to $x$ and $y$ respectively and placing the result in $z$.
It is defined as
\begin{itemize}
\item $M_{zxycc'}= 1$, if $z = x+y+c'\ [2]$ and $c = \lfloor (x+y+ c') / 2\rfloor$,
\item $M_{zxycc'}= 0$ otherwise.
\end{itemize}
and its graphical representation is,
\begin{center}
\includegraphics[scale=0.3]{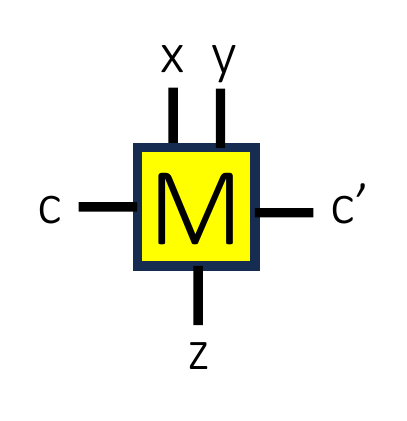}
\end{center}
Second, we construct the shift MPS defined as,
\begin{equation}
t_\alpha = \tfrac{1}{6}\left( -\delta_{\alpha,-1} + 4 \delta_{\alpha,0} 
+ 4 \delta_{\alpha,+1} - \delta_{\alpha,+2} \right).
\end{equation}
Since it is a sum of four rank-$1$ terms, it is at most of rank $4$ (actually, in a minor 
optimization of this construction that we omit for clarity, the terms $4\delta_{\alpha,0}$
would be omitted and the rank would be at most $3$). Its graphical representation reads,
\begin{center}
\includegraphics[scale=0.3]{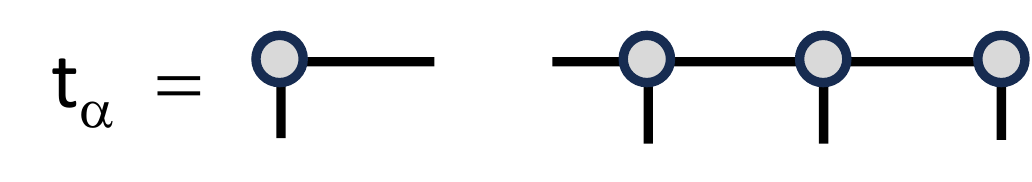}
\end{center}
Now the MPS of the interpolant $g_\alpha$ is obtained by contracting the following network,
\begin{center}
\includegraphics[scale=0.3]{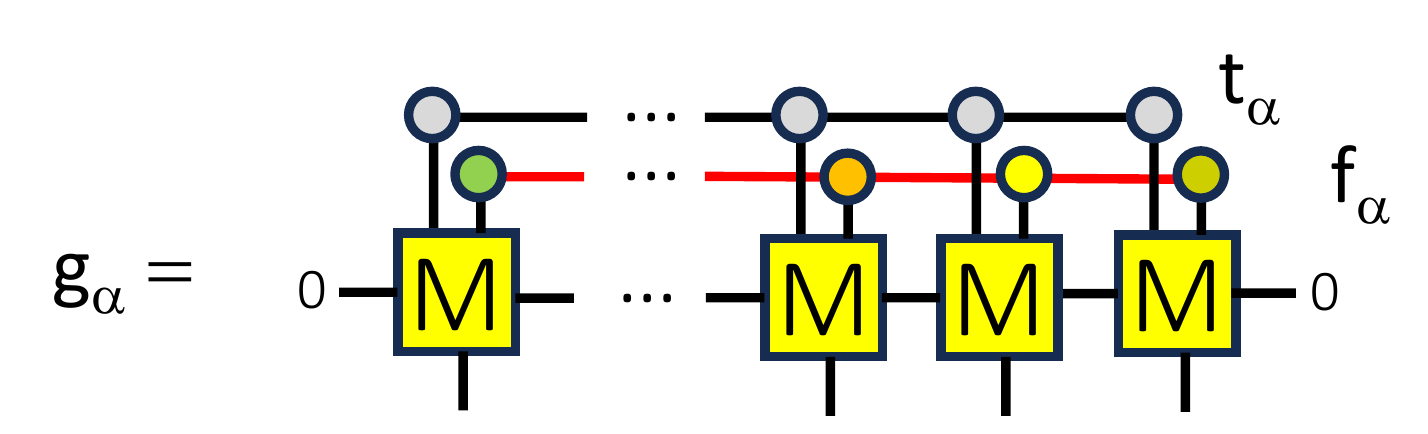}
\end{center}
(we refer to \cite{Waintal2026} for the proof).

\bibliography{PDE}

\end{document}